\documentclass[hyper, notoc]{JHEP3} 
\usepackage{epsfig} 

\newcommand{\tev}{\ {\rm TeV}}
\newcommand{\gev}{\ {\rm GeV}}
\newcommand{\be}{\begin{equation}}
\newcommand{\ee}{\end{equation}}
\newcommand{\bea}{\begin{eqnarray}}   
\newcommand{\eea}{\end{eqnarray}}

\def\simlt{\stackrel{<}{{}_\sim}}

\title{Perturbative, Non-Supersymmetric Completions of the Little
Higgs} \author{Puneet Batra \\ High Energy Physics Division, Argonne
National Lab, Argonne, IL 60439} \author{David E. Kaplan\\ Department
of Physics and Astronomy, Johns Hopkins University, Baltimore, MD
21218} \abstract{The little Higgs mechanism produces a light 100 GeV
Higgs while raising the natural cutoff from 1 TeV to 10 TeV.  We
attempt an iterative little Higgs mechanism to produce multiple
factors of 10 between the cutoff and the 100 GeV Higgs mass in a
perturbative theory.  In the renormalizable sector of the theory, all
quantum corrections to the Higgs mass proportional to mass scales
greater than 1 TeV are absent -- this includes quadratically
divergent, log-divergent, and finite loops at all orders.  However,
even loops proportional to scales just a factor of 10 above the Higgs
(or any other scalar) mass come with large numerical factors that
reintroduce fine-tuning.  Top loops, for example, produce an expansion
parameter of not $1/4\pi$ but $1/5$.  The geometric increase in the
number of fields at higher energies simply exacerbates this
problem. We build a complete two-stage model up to 100 TeV, show that
direct sensitivity of the electroweak scale to the cutoff is erased,
and estimate the tuning due to large numerical factors.  We then
discuss the possibility, in a toy model with only scalar and gauge
fields, of generating a tower of little Higgs theories and show that
the theory quickly becomes a large-$N$ gauge theory with $\sim N$
fundamental scalars.  We find evidence that at least this toy model
could successfully generate light scalars with an exponentially large
cutoff in the absence of supersymmetry or strong dynamics.  The
fine-tuning is not completely eliminated, but evidence suggests that
this result is model dependent.  We then speculate as to how one might
marry a working tower of fields of this type at high scales to a
realistic theory at the weak scale.  }
\preprint{ANL-HEP-PR-04-139}

\begin{document}

%%%%%%%%%%%%%%%%%%%%%%%%%%%%%%%%%%%%%%
\section{Introduction} %%%%%%%%%%%%%%%%%%%%%%%%%%
\label{sec:intro} %%%%%%%%%%%%%%%%%%%%%%%%%%%%%%%
%%%%%%%%%%%%%%%%%%%%%%%%%%%%%%%%%%%%%%

For over a quarter of a century, the gauge hierarchy problem -- the
exponential size of the Planck scale to weak scale ratio and the
quantum instability of this ratio -- 
has dominated our guidance in the
search for physics beyond the Standard Model.  Remarkably, only a
handful of viable possibilities have even been proposed.  Even more
remarkably, only one - softly broken supersymmetry - is weakly
coupled.

A more experimentally driven version of the hierarchy problem would be
the fact that the natural (in the 't Hooft sense \cite{'tHooft:1979bh}) 
scale of new physics (500 GeV - 1 TeV) is lower than the scales indirectly
probed by electroweak precision measurements (5-10
TeV).  This ``weak'' version of the problem has been coined the
``little hierarchy problem'' \cite{mt}.
A model of
electroweak physics that is weakly coupled (such as supersymmetry) has
much greater control over contributions to electroweak precision
parameters, both in the ability to calculate and suppress. Thus
supersymmetry has the upper hand.  However, one must note that large
$N$ gauge theories with large 't Hooft coupling appear to have a
weakly coupled dual (except for threshold effects/boundary terms) 
\cite{Maldacena:1997re} in Randall-Sundrum models \cite{Randall:1999vf} and
could garner some of the same flexibility \cite{Agashe:2002bx}.

A recently discovered class of models, ``little Higgs'' theories \cite{Arkani-Hamed:2001nc}-\cite{Schmaltz:2004de},
involve a Higgs boson as a pseudo-Nambu-Goldstone boson (pNGB) with
additional structure that allows the cutoff to live naturally at
around 10 TeV.  The extra order of magnitude comes from the collective
nature of explicit symmetry breaking in the Lagrangian, or ``collective symmetry breaking''.  
No one operator is enough to render the Higgs a {\it pseudo}-NGB and thus no
quadratically divergent contributions to the Higgs mass appear at one
loop.  The theory is then weakly coupled up to 10 TeV and can control
electroweak precision contributions \cite{Peskin:1990zt,Csaki:2002qg,Hewett:2002px}
 via T-parity \cite{Cheng:2003ju,Cheng:2004yc} or
differing global symmetry vacuum expectation values (vev's) \cite{Kaplan:2003uc,Csaki:2003si}. 

Ultraviolet (UV) completions of little Higgs theories are needed to
help justify the explicit symmetry breaking structure that keeps the
Higgs light. A natural UV completion for most little Higgs
theories is that of the composite Higgs scenario \cite{Kaplan:1983fs}
which is a technicolor-like theory.  This matches a theory that solves
the big hierarchy problem (new strong dynamics) with one which solves
the little hierarchy problem (collective symmetry breaking).  While
electroweak precision contributions can now be suppressed (and better
estimated) in a class of technicolor theories, the actual model
building remains difficult and requires dynamical assumptions.  See,
for (the only) example, Ref. \cite{Katz:2003sn}.  (However, see also
recent work generating composite Higgs models from a Randall-Sundrum
type UV completion \cite{Agashe:2004rs,Contino:2003ve}).

The little Higgs, however, offers a completely new possibility for UV
physics.  The symmetry breaking which produces the pNGB Higgs in the
first place could be generated by a second
 weakly coupled scalar.  Its mass
($\sim 1$ TeV) could be naturally light with a cutoff of 100 TeV if
the second scalar had its {\it own} little Higgs mechanism.  This would be a weakly
coupled theory up to 100 TeV with a scalar (the Higgs) of mass roughly
100 GeV, and naturally.  The breaking scale which generated the
``upper'' pNGB could then in principle be generated by another
fundamental scalar with its own little Higgs mechanism, etc.  The idea
would be a cascade of symmetry breaking from the Planck scale
downward to the weak scale, all weakly coupled and all without
supersymmetry. 

An iterative approach, where each stage can be constructed from the
lower stage with a simple prescription, presents the only realistic
way of producing such a tower. Each stage would be characterized by a
symmetry breaking scale ($f_i$) which produces the naturally lighter
(masses of order $f_{i-1} \sim f_i/4\pi$) pNGB of the stage underneath
it. The obstacles toward producing such a tower are significant.  The difficulties
come in two types:  structural -- reducing the sensitivity of light scalars to higher
mass scales -- and cumulative -- avoiding large numerical factors due to large 
numbers of fields from the additional structure.

The structural concerns are the most obvious of issues:  how does one suppress
scalar mass sensitivity to higher scales?  Collective symmetry breaking
normally eliminates one-loop quadratic divergences, but not those at two loops.  
Making the cutoff higher
would make these smaller contributions important and destabilize the
weak scale.  In addition, there are always finite corrections which appear at
one loop.  Because from the start the theory explicitly breaks all of
the symmetries which could keep the Higgs an exact NGB, one must worry
about one-loop finite corrections proportional to higher scales.

We have found a structure which solves 
the above sensitivity issues in terms of the parametric dependence on
higher scales.  The model is of the type in Ref. 
\cite{Kaplan:2003uc}, the ``simple group'' little Higgs, where the gauge group is
of the form $SU(N)\times U(1)$ with a set of scalars in the
fundamental representation, $\phi^{}_i$.  The UV extensions at each stage
are {\it linear}-sigma models (as opposed to the non-linear sigma models typically
used in Little Higgs theories) and thus are weakly coupled.  The key is that at every stage
there are additional $U(1)$ global symmetries, similar to chiral
symmetries for fermions, which protect every breaking scale.  These
scales are generated by mass terms that mix different $\phi^{}$ fields,
namely $b_{ij} \phi_i^\dagger \phi^{}_j$, and therefore break $U(1)$
symmetries associated with each field.  These ``chiral'' symmetries
remove the dangerous radiative effects.  They remove quadratically 
divergent contributions to symmetry breaking scales at {\it all} loops
by allowing only dimension-4 $SU(N)$-breaking operators.  In addition,
they control the scale of the finite and log-divergent loops by requiring a dimensionful
parameter $b_{ij}$ to appear in the loops.  The quartic structure is like that in
\cite{Kaplan:2003uc} and is repeatable for each stage, and thus tree-level quartics
exist for every symmetry breaking scale, again without destabilizing the structure.
Thus, amazingly enough, we have shown that the scalars of the scale $f_k$ are only sensitive to $f_{k+1}/4\pi$ and never to $f_{k+2}$ or $f_{k+3}$.

The inherent assumption in the above discussion is that the expansion (loop) parameter is
$1/4\pi$.  This is in fact not the case in the standard model, and even less true once the extra
fields and structure needed for collective symmetry breaking is included.
Contributions to the Higgs mass are (at least) proportional to $\sim 6 y_t^2 ({\tilde f}/4\pi)^2$, 
where ${\tilde f}$ is the mass scale of new 
partner states and $y_t \sim 1$ is the Yukawa coupling.  Thus instead of
 $f_0/f_1 \simeq 1/4\pi$, the expansion parameter is at least $f_0/f_1 \simeq 1/5$.  
This occurs in every little Higgs model
and thus a natural size for $f_1$ is typically $\sim 700 - 800 \gev$.  When the theory
is UV completed at strong coupling, the top Yukawa and other symmetry-violating couplings are
small spurions and don't affect the estimate\footnote{If one includes the 
number of Goldstone flavors \cite{Chivukula:1992nw}, the estimate for the cutoff reduces by a 
factor of 2 or so.  Putting the cutoff at the scale of unitarity breakdown reduces it by a factor of 3
or so \cite{Chang:2003vs}.  However, an unknown parameter of order unity accompanies these
estimates.} for the cutoff $\Lambda\simeq 4\pi f$.  However, in the weakly coupled option, a linear
sigma model, the breaking scale $f$ becomes sensitive to the same loop corrections 
that affect the electroweak vev ({\it i.e.}, the top loop).  In addition, there are more fields in order
to produce collective symmetry breaking in every sector and therefore that $1/5$ is multiplied by
a factor of at least two.  The hierarchy between scales wants to collapse
under the weight of the field content, unless one fine-tunes at every stage.

Any iterative tower will have a geometrically increasing number of fields,
which will destabilize the hierarchy discussed above with
coefficients proportional to the size of the gauge group or number of fields, 
making even ``parametrically acceptable'' corrections too large.  Eventually one would
even expect the theory to become strongly coupled.  In addition, the gauge groups of 
hypercharge and color are asymptotically unfree due to the large numbers of fields and 
quickly hit a Landau pole in the (nearby) UV. 

However, an amazing thing happens when you pare down the theory to its bare minimum,
{\it i.e.}, without color, hypercharge or even fermions.
The theory quickly becomes a large
$N$ gauge theory with $N_s \sim 3N$ scalars in the fundamental representation.
The gauge coupling remains asymptotically free in the theory and scales in the
running as $1/N$.  In fact, if the scales are fixed at $4\pi$ apart from each other, 
the 't Hooft coupling $g^2 N$ {\it remains perturbative to arbitrarily
high scales}.  Even more importantly, there exist approximate tracking solutions 
to the renormalization group equations such that the quartic couplings also run asymptotically
free, or better yet, they track the gauge coupling from the UV (with values of order $1/N$)
into the IR (with values of order unity).  In addition, there is a consistent structure which guarantees
the correct vacuum (mis)alignment at each stage.  The only problem left with this model is the
actual coefficients in the beta functions -- the asymptotic average value of the 't Hooft coupling
in our model is $\alpha_{'t Hooft}/4\pi \equiv g^2 N/16\pi^2 \sim 1/4\pi$ and not $\sim 1/16\pi^2$
and therefore to maintain the $4 \pi$ scaling between breaking scales would require some tuning
at each stage.  We take this to be a failure of the specific model and not of the overall setup and 
speculate what steps could be taken to build a ``free-standing'' tower.  Such a tower would be the
first example of a weakly coupled tumbling gauge theory \cite{Raby:1979my}, and thus perhaps the only reliable one.

In this paper we present a complete model of electroweak symmetry breaking up to $100 \tev$
which is both weakly coupled and non-supersymmetric.  We include the full gauge structure
of the standard model and all Yukawa couplings.  We also carefully estimate all contributions
to scalar masses and outline all sources of fine tuning.  It is exactly the large numerical factors
described above that hurt this model.  We then describe a preliminary toy model which exhibits 
some of the large $N$ behavior necessary to maintain a stable tower.  We then speculate as to
how one might connect such a tower to a realistic description of electroweak physics without
introducing too much fine tuning.

Recently as a first attempt, something to the effect of a two-stage
little Higgs model has been produced \cite{Kaplan:2004cr}.  It provides a weakly
coupled theory of electroweak symmetry breaking up to about 30 TeV.
In that model, an $SU(3)\times U(1)$ \cite{Schmaltz:2004de} model  was completed
by two copies of the ``littlest Higgs''.  Because the quartic of the
bottom theory is generated radiatively, the prediction for the
breaking scale of the electroweak symmetry is naively that of the
breaking scale of the $SU(3)$ group as well and therefore requires
a ten percent fine-tuning to generate the electroweak scale.  The main 
result of that paper was to give an alternative to T-parity (separating the
new physics scales of the top and gauge sectors) for suppressing contributions
to precision electroweak observables.  The lack of a tree-level quartic at the lowest
stage, and the fact that the two stages are quite dissimilar, imply that such a model
does not have the potential for a repeatable structure.  Additional interesting possibilities
involve non-supersymmetric theories which flow close to a non-trivial and supersymmetric
fixed point \cite{Goh:2003yr,Strassler:2003ht} and field theory orbifolds of supersymmetric
theories \cite{Frampton:1999yb} (though in the latter the hierarchy is destabilized by
corrections subleading in $N$ \cite{Csaki:1999uy}).

In Section \ref{sec:10} we describe the lowest layer of this
arrangement, a linear sigma model completion of the $SU(4)$ Simple
little Higgs \cite{Kaplan:2003uc}. This layer is cutoff at $10 \tev$ 
where in the non-linear sigma model the theory would become strongly coupled.
For us, $10 \tev$ (or $\sim 4\pi \times 1 \tev$) is the na\"{i}ve
cutoff required to maintain a natural theory with collective symmetry breaking.  
We show that some of the fine-tuning is reintroduced due to the multiplicity of
fields.  In Section \ref{sec:100} we lift the cutoff to
$100 \tev$ and fill the new energy regime
with a spontaneously broken $SU(12)$ little Higgs theory.  We explain both
the insensitivity of the Higgs mass to the high cutoff and the (again) partial reintroduction
of fine-tuning due to the number of fields.  In the final section, we discuss the possibility 
of stacking theories with collective symmetry breaking 
and show preliminary results for a toy model with partial
success.  We argue that the existence of a full working tower is plausible and 
describe possible paths for getting there.  In the appendices we describe some 
of the detailed structure of the symmetry breaking and also give the renormalization 
group equations for the couplings in the theory valid at any stage.

%%%%%%%%%%%%%%%%%%%%%%%%%%%%%%%%%%%%%%
%%%%%%%%%%%%%%%%%%%%%%%%%%%%%%%%%%%%%%
\section{Below 10 TeV} %%%%%%%%%%%%%%%%%%%%%%%%%%
\label{sec:10} %%%%%%%%%%%%%%%%%%%%%%%%%%%%%%%
%%%%%%%%%%%%%%%%%%%%%%%%%%%%%%%%%%%%%%

\FIGURE[t]{
\label{fig:10}
\epsfig{file=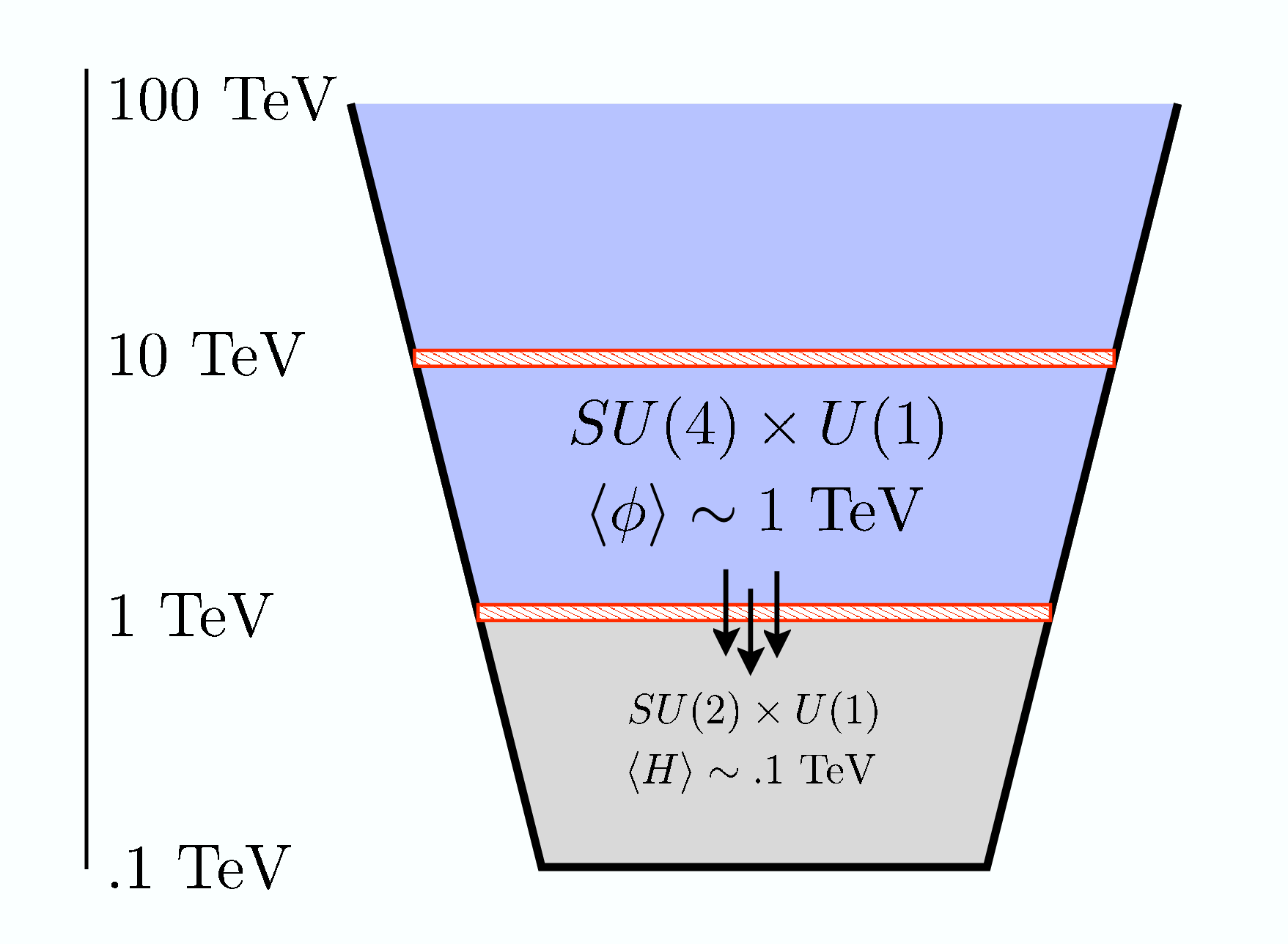,width=3.5 in}
\caption{The energy scale and dependence of just a single
stage of a Little Higgs theory. The tower of effective field theories (EFTs)
begins with an $SU(4) \times U(1)$ EFT that spontaneously breaks at $1
\tev$, leaving behind an $SU(2)_W \times U(1)$ EFT which spontaneously
breaks at $.1 \tev$.}  }

Below 10 TeV the model is a linear-sigma model version of the
$SU(4)\times U(1)_X$ Little Higgs \cite{Kaplan:2003uc}: At $f_1 \sim 1
\tev$ a set of scalars, $\phi^{}$, collectively and spontaneously
breaks an $SU(4)\times U(1)_X$ gauge symmetry to the electroweak gauge
group, $SU(2)_W \times U(1)_Y$. Two $SU(2)_W$ doublets, $H_1 \ \& \
H_2$, are pNGB, and they remain in the low energy spectrum. These
Higgs fields have tree-level interactions that promote $SU(2)_W \times U(1)_Y$
breaking to $U(1)_{{\rm E \& M}}$. We summarize the fields, scales,
and symmetry structure in Figure \ref{fig:10}.

The novelty of Little Higgs theories lies in the separation of scales.
While $SU(4)\times U(1)_X$ breaks at $f_1 \sim 1 \tev$, $SU(2)_W
\times U(1)_Y$ breaks at a lower scale, $f_0 \sim .1 \tev$. This
hierarchy between $f_1$ and $f_0$ is enforced  in the $SU(4) \times 
U(1)_X$ gauge theory by a set of
global symmetries and their spurion structure. These approximate
symmetries ensure cancellations of all quadratically divergent loop
diagrams from renormalizable operators that might
contribute to the masses of the $H_1 \ \& \ H_2$ and destabilize the
ratio $f_0/f_1$.

For just this section we impose a cutoff of $\Lambda = 10 \tev $ on
the theory. Without additional UV structure, naturalness would be lost
if we allowed $\Lambda$ to be greater than $10 \tev$--although $H_1 \
\& \ H_2$ are protected from quadratic divergences, the $\phi^{}$ fields
are not. In the non-linear sigma-models already present in the
literature, \cite{Arkani-Hamed:2001nc}-\cite{Schmaltz:2004de}, 
$10 \tev$ corresponds to the
scale of strong-coupling due to the absence of the symmetry-breaking
radial modes. By working in a linear-sigma model, we keep the theory
at weak-coupling near $10 \tev$, which allows us to extend the theory
perturbatively to $100 \tev$.  In Section \ref{sec:100}, we actually
lift $\Lambda$ to $100 \tev$ and detail the perturbative UV structure
that justifies our chosen spurion structure and maintains an insensitivity of 
the Higgs mass to high scales.

%%%%%%%%%%%%%%%%%%%%%%%%%%%%%%%%%%%%%%
\subsection{$SU(4) \times U(1)_X$ Breaking} %%%%%%%%%%%%%%%
\label{sec:su4} %%%%%%%%%%%%%%%%%%%%%%%%%%%%%%

 The scalar sector of the $SU(4) \times U(1)_X$ gauge theory consists
of fundamental fields, $\phi^{}$, with the same $U(1)_X$ charge, $-1/4$. These
$\phi^{}$ fields are arranged in pairs, $\phi^{}\ \& \ \phi'$, with
interactions designed to produce Goldstone bosons; we visually depict
these interactions in Figure \ref{fig:2phi}. This structure is
familiar -- if we replace $\phi^{} \ \& \ \phi'$ with $H_1 \ \& \ H_2$
we find the usual scalar sector of a standard 2-Higgs-Doublet-Model
(2HDM).

\FIGURE{
\epsfig{file=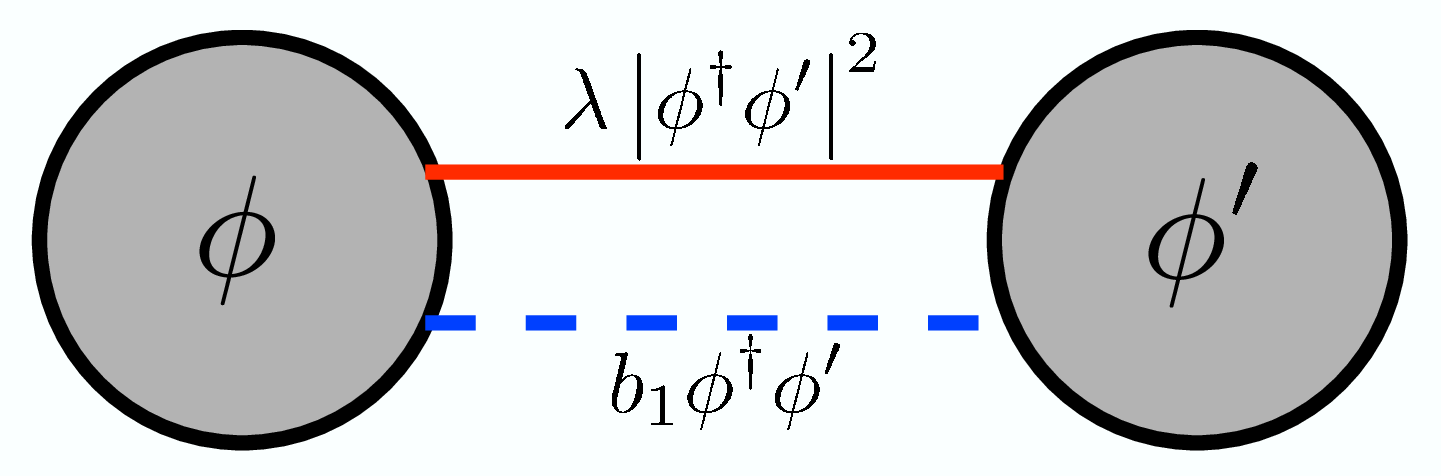,width=2.5in}\caption{}\label{fig:2phi}}As
visual shorthand Figure \ref{fig:2phi} has two elements: shaded circles (which always represent fundamental scalar fields that have vev's) and lines (which represent interactions between fields). Specifically, the diagram in Figure \ref{fig:2phi} represents the potential \be V
\supset -b_1 \left(\phi^{\dagger} \phi' + {\rm h.c.}\right) +m_1^2 \left( \left| \phi^{} \right|
^2 + \left| \phi' \right| ^2\right)+ \lambda \left| \phi^{\dagger}
\phi' \right|^2.
\label{eq:2}
\ee For simplicity we've assumed that all the scalar masses in the
theory are identical, though none of our results depend on this
degeneracy. We also set all dimensionful operators to be of order $
\sim 1 \tev$, the seemingly natural scale given a cutoff of $10 \tev$ 
(we discuss this assumption further in Section
\ref{sec:10qc}).

In the potential of Equation \ref{eq:2}, the $b_1$-term promotes
spontaneous symmetry breaking while the quartic $\lambda$ provides
stability around the minimum. The potential has a global $U(4)$
symmetry, under which $\phi^{} \ \& \ \phi'$ are fundamentals. Due to
this structure, $\phi^{} \ \& \ \phi'$ acquire vev's $\sim 1 \tev$ \be
\langle \phi^{} \rangle = \left(\begin{array}{c}0 \\ 0 \\ 0 \\ f_1
\end{array}\right)\ \ \ \langle \phi' \rangle =
\left(\begin{array}{c}0 \\ 0 \\ 0 \\ f_1 \end{array}\right), {\ \rm
with}\ f_1 = \frac{b_1-m_1^2}{\lambda}.  \ee The vev's break the
global $U(4)$ symmetry to $U(3)$ and produce
7 would-be Goldstone bosons. The would-be Goldstone bosons are eaten by the
gauge bosons as the vev's also break the gauge symmetry down to
$SU(3)\times U(1)$.

\FIGURE[t]{
\epsfig{file=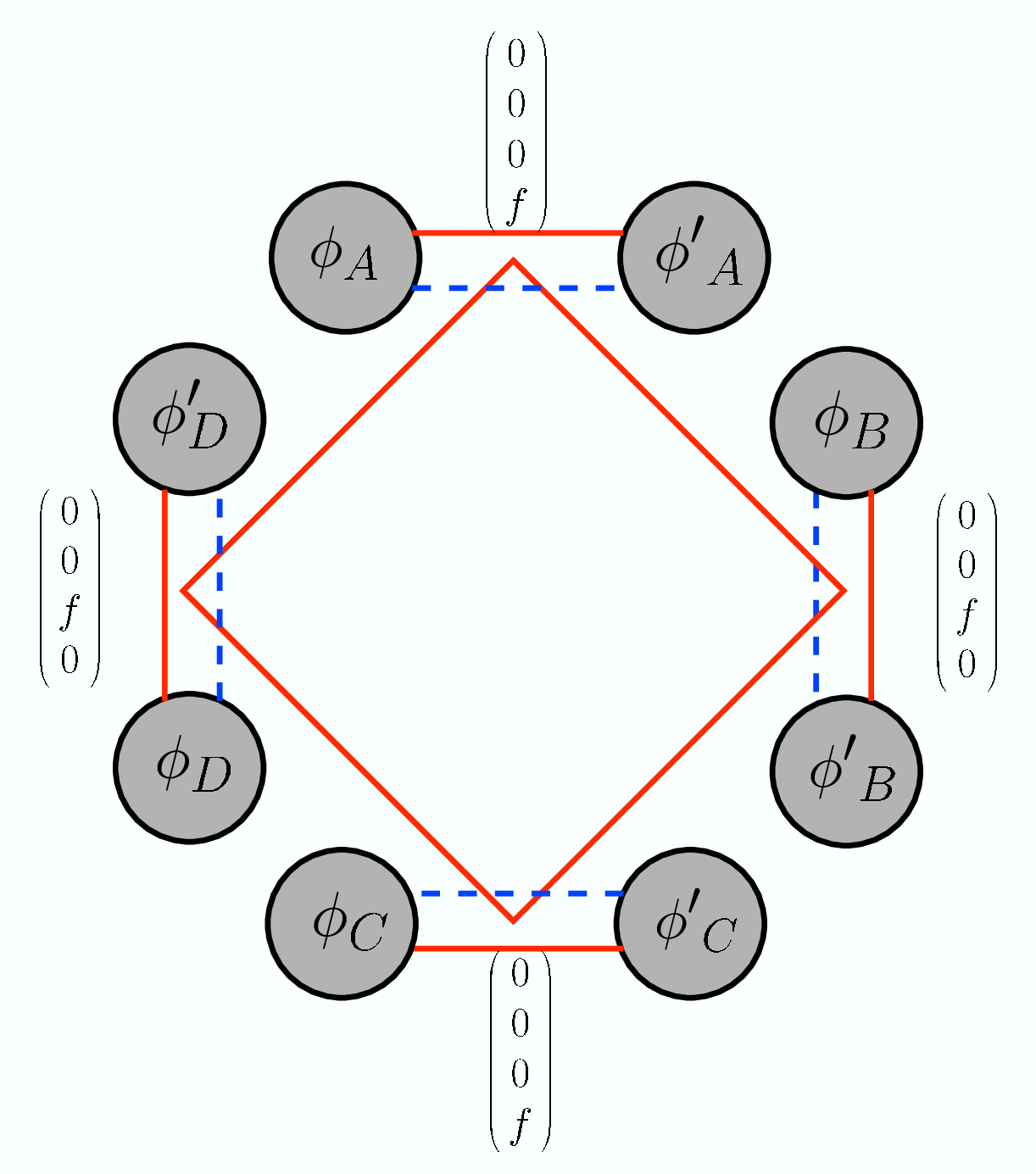,width=3.5in}
\caption{All 8 fields which contribute to the $SU(4)$ breaking. The
visual shorthand is that of Figure \ref{fig:2phi} with one
modification.  The largest quartic
links (those connecting two sets of pairs) have ends in the middle of
each pair---which is shorthand for four separate quartics (equivalent to a red,
solid line between each element of one pair and each element of the
other pair).  These specific quartic operators are described in
Equation \ref{eq:quartic}. }
\label{fig:8}
}

The full $SU(4) \times U(1)_X$ gauge theory contains four
such pairs---each with the interactions described in Equation
\ref{eq:2}. Each of these pairs in isolation
(no interactions between pairs, no gauge interactions) produces a set
of 7 Goldstone bosons which includes one $SU(2)$ doublet, for a total
of four Goldstone $SU(2)$ doublets.  
After we turn on interactions between pairs and gauge the $SU(4)$
symmetry, these four doublets divide into two groups: two of the four
doublets are eaten by the $SU(4)/SU(2)_W$ gauge bosons; the remaining
two are the pNGB we desire, $H_1 \ \& \
H_2$.  They have charges $(2, -1/2)$ under the remaining
$SU(2)_W \times U(1)_Y$ EFT and are massless at tree-level.

We show the additional interactions between the pairs in Figure
\ref{fig:8}, where we arrange the pairs in a square whose edges
represent quartic interactions between fields. With the labels of
Figure \ref{fig:8} these additional quartic interactions are 
\be V
\supset \lambda \left( \left| \phi_A^{\dagger} \phi^{}_B \right|^2
+\left| {\phi'}_{A}^{\dagger} \phi^{}_B \right|^2+\left|
\phi_A^{\dagger} \phi'_{B} \right|^2 +\left| {\phi'}_A^{\dagger}
\phi'_B \right|^2 \right).  
\label{eq:quartic}\ee 
Similar terms exist only for fields which are at neighboring vertices
on the square, $(A,B) \rightarrow (B,C), (C,D)$ or $(D,A)$, for a
total of sixteen new quartic interactions.  The additional quartic
interactions stabilize the $SU(4) \times U(1)_X \rightarrow SU(2)_W
\times U(1)_Y$ breaking pattern by pushing the pairs of vev's (at
tree-level) into the structure indicated in Figure \ref{fig:8}.  The
(mis)alignment of the vev's which break $SU(4)$ can be guaranteed by
the combination of the positive quartics between adjacent sites and
the introduction of small $b$ terms and quartic couplings which stretches
across the diamond to connect sites at opposite ends.  A more detailed
analysis of the remainder of the symmetry breaking structure is
provided for reference in Appendix \ref{app:su4}.

Although each pair of fields of Figure \ref{fig:8} has a global $U(4)$
symmetry, the $SU(4)^4$ subgroup is explicitly broken to the diagonal
$SU(4)$ by the extra quartic interactions between pairs. This diagonal $SU(4)$ is
actually the gauged symmetry, and the gauge interactions also
explicitly break the $SU(4)^4$ to the diagonal. With all
interactions turned on, the global symmetry is reduced from $U(4)^4$ to $SU(4)
\times U(1)^4$.

As shown in Ref. \cite{Kaplan:2003uc}, after the $SU(4) \times U(1)_X$ 
symmetry breaks, the pNGB $H_1 \ \&\  H_2$ are left with their
own tree-level quartic interaction: $ V \supset \lambda
\left|H_1^{\dagger} H_2\right|^2$.  Figure \ref{fig:8to2} shows the
final Higgs structure. It is important to note that the generated
quartic is the same size as the $SU(4) \times U(1)_X$ quartics of
Equation \ref{eq:quartic}. This quartic interaction is generated at
tree-level after integrating out singlets with trilinear interactions
and masses of order $f_1$. 

We can generate the $b_0 \sim .1 \tev$ term
that's required for $SU(2)_W \times U(1)_Y$ breaking by including seed
terms, e.g. $b_0 \phi_A^{\dagger} \phi^{}_B, \ldots$, in the original
$\phi^{}$ theory. These are the only terms that violate the $U(1)^4$
symmetry, so they are technically natural. Of the four would-be Goldstone 
electroweak singlets from the $U(1)^4_{\phi}$ breaking, two singlets are 
eaten by gauge bosons and the other two singlets pick up weak scale 
masses from the $b_0$ terms. The phenomenology of these singlets is 
described in Ref. \cite{Kilian:2004pp}, and otherwise the phenomenology
is that described in \cite{Kaplan:2003uc}.  For example, the ratio of Higgs
vev's ({\it i.e.}, $\tan\beta$) goes like the inverse ratio of their diagonal masses.
The top loop suppresses the ``up-type'' mass and therefore prefers large $\tan\beta$, 
which is good for keeping down the Yukawa coupling.

  \FIGURE[t]{\epsfig{file=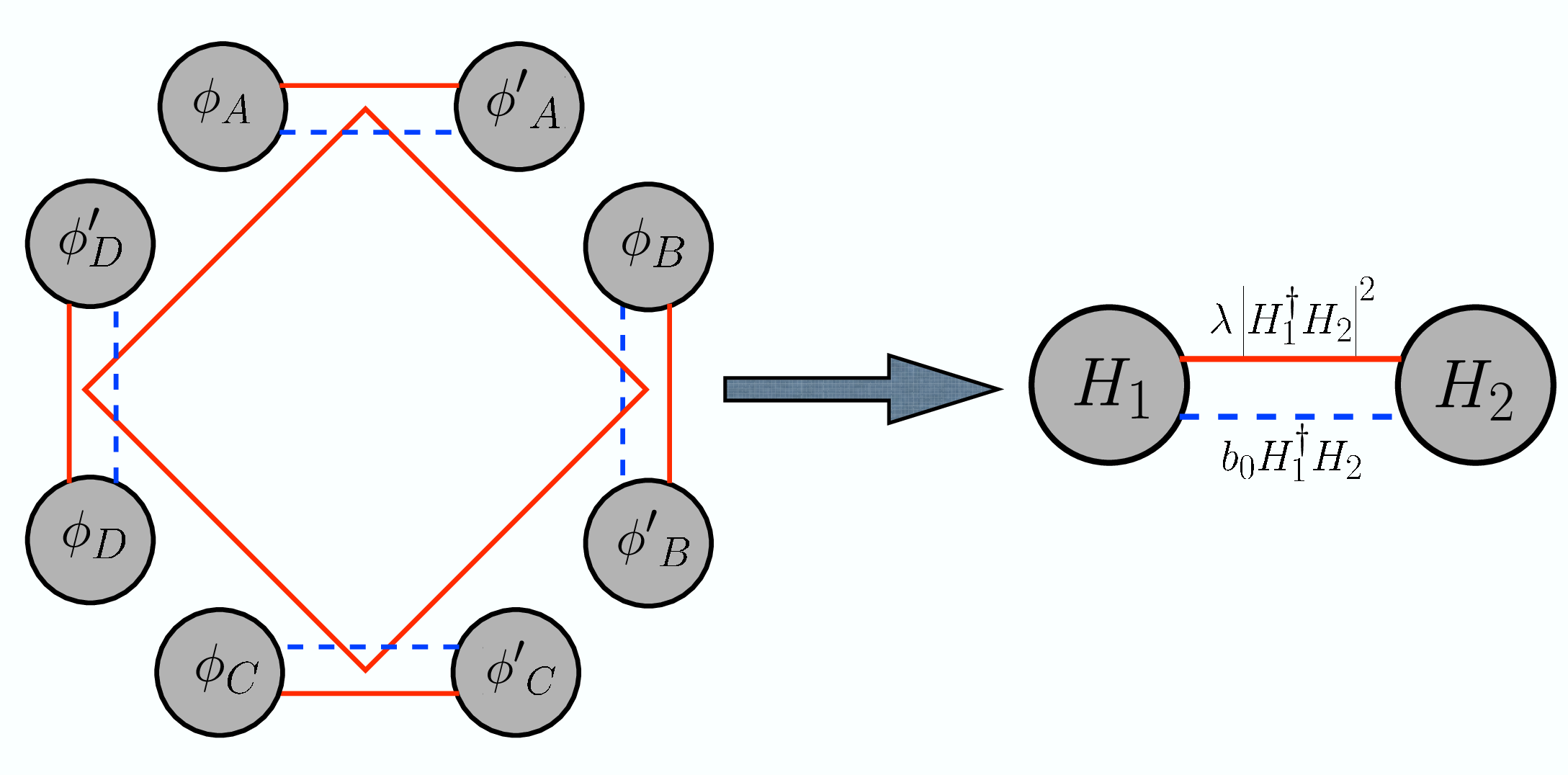, width=5in}\caption{The full structure
of symmetry breaking below $10 \tev$. The 8 fields of the $SU(4)$
Little Higgs theory spontaneously break and leave behind the indicated
2HDM.}\label{fig:8to2}}
%%%%%%%%%%%%%%%%%%%%%%%%%%%%%%%%%%%%%%
\subsection{Fermions \& Yukawa Interactions}%%%%%%%%%%%%%%%%
\label{sec:u1x} %%%%%%%%%%%%%%%%%%%%%%%%%%%%%%%

Fermions must also be embedded in representations of $SU(4)$, with
$SU(4) \times U(1)_X$ gauge invariant Yukawa interactions. For example, the third generation quark doublet becomes a fundamental of $SU(4)$, with $\left(SU(3)_c, SU(4), U(1)_X\right)$ quantum numbers $\left(3, 4, 5/12\right)$.  We define
$\hat{Q}_{3}=\left(Q_3, T_1, T_2\right)$ with $Q_3$ the third generation $SU(2)_W$ quark doublet. We write down a Yukawa interaction with
 \be V \supset  \left\{y^{}_{TA} \phi_A^{\dagger} T^c_A +y^{}_{TB}{\phi}_B^{\dagger}{T}^c_B + y^{}_{TC} \phi_C^{\dagger} T^c_C \right\} \hat{Q}_{3},  \ee where $T^c_A, {T}^c_B \ \& \ T^c_C$ have
quantum numbers $\left( \overline{3}, 1,
-2/3\right)$. These interactions also preserve the global $SU(4) \times
U(1)^4$ symmetry, just like the gauge and scalar interactions.\footnote{Note that there is substantial freedom in choosing which 3 $\phi$ fields among the 8 total are listed in the Yukawa interaction.}

In the $SU(2)_W$ EFT below $\langle \phi^{} \rangle$, the above Yukawa
structure reproduces the SM fields and interactions. Consider the
limit of degenerate $y^{}_T$, with $y^{}_{TA} = y^{}_{TB} = y^{}_{TC}$, which
gives particularly simple results:

\bea
V &\supset& y_t H_1^{\dagger} Q t^c + m_1 T_1 T_1^c +m_2 T_2 T^c_2,
\label{eq:tope}\eea
with $y_t = y_T/\sqrt{2}$, $m_1 = \sqrt{2} y_T f_1,$ and $ m_2 = y_T f_1$. The diagonalized fields are  $t^c = 1/\sqrt{2}\left( T^c_A -T^c_C\right)$, $T_1^c =1/\sqrt{2}\left( T^c_A + T^c_C\right)$, and $T_2^c = T_B^c$. We discuss below the more general situation where the $y_{TX}$ (and $f_1$) differ.

We add a mass for the bottom quark with a non-renormalizable Yukawa
interaction: 
\be
\frac{y_B}{\Lambda^2} \phi^{}_A \phi^{}_B {\phi}^{}_{D} \hat{Q}_{3} b^c,
\ee 
where the $SU(4)$ indices are tied together with an antisymmetric 4-index
epsilon tensor. After $SU(4)$ symmetry breaking, $H_2$ plays the role of the down-type Higgs, with Yukawa $y_b= y_B*f_1^2/\Lambda^2$, so $y_B \sim 1$. This operator preserves the global $SU(4) \times U(1)^4$ symmetry.

Anomaly cancellation occurs between the three
generations \cite{Kong:2003ii,Kaplan:2004cr,Schmaltz:2004de}. The first two
generation of quarks, $\hat{Q}_{1,2}$, are embedded as $\left(3,
\overline{4}, -1/12\right)$. Therefore, the {\it down}-type fermions have
renormalizable interactions, while the up-type quarks have
non-renormalizable interactions. The first two generations of down-type fermions have interactions like
\be V \supset \left\{y_{DB} {\phi'}^{}_B D^c_B +y_{DC} {\phi^{}}_C {D}^c_C
+ y_{DD} {\phi'}^{}_D D^c_D\right\} \hat{Q_1}, \ee where $D^c_B, {D}^c_C \
\& \ D^c_D$ are $\left(\overline{3},1, 1/3\right)$.
The first two generations of up-type fermions get masses from the interactions 
\be 
\frac{y_U}{\Lambda^2} \phi_C^{\dagger} {\phi'}_B^{\dagger} {\phi}_D^{\dagger}
\hat{Q}_{1} u^c,  \ee 
with $u^c=\left(\overline{3}, 1, -2/3\right)$.

For the first two generations, we are forced to choose non-degenerate Yukawa
couplings. In fact, to ensure that the heavy lepton partners are heavy
enough, we must take $y_{DC}, y_{DD} \sim 1$, while $y_D$ has size
similar to the relevant Yukawa coupling in the SM. This disparity in
couplings leads to mixing between the light and heavy fermions after
electroweak symmetry breaking. However, the effect of this mixing is
minimized in the precision electroweak measurements when we begin to consider
non-degenerate ratios of vev's (e.g. $f_A/f_C \sim 3-5$) \cite{Kaplan:2003uc,Csaki:2003si}.

All three generations of leptons are embedded as $(1, 4,
-1/4)$. For each generation, three ($SU(4)\
{\rm and} \ U(1)$) singlet fields $N$ field with charges $\left( 1, 1, 0\right)$ are used (analogous to the
$D^c$ and $T^c$ above), while
one $e^c$ field with charges $\left(1, 1, -1\right)$ is introduced to give the electron a mass (analogous
to the $u^c$ and $b^c$ above).

%%%%%%%%%%%%%%%%%%%%%%%%%%%%%%%%%%%%%%
\subsection{Loop Corrections}%%%%%%%%%%%%%%%%%%%%%%%
\label{sec:10qc}%%%%%%%%%%%%%%%%%%%%%%%%%%%%%%

Although $H_1 \ \& \ H_2$ remain massless at tree-level, no unbroken
symmetry prevents masses from being generated at loop level. The Higgs
sector is massless in the limit of the exact unbroken $U(4)^4$
symmetry, but we have dimensionless ${\cal O}(1)$ spurions ($g, y_T,
\lambda$) which explicitly break the global symmetry to $SU(4) \times
U(1)^4$.  We estimate the size of the Higgs masses, and the naturalness of
the theory, by calculating the size of the loop contributions to
operators that we've ignored in the theory---assuming that physics
above the cutoff contributes corrections of the same magnitude. 

We find that these loops are parametrically under control, since the
unbroken global $U(1)$ symmetries prevent dangerous dimension-two
operators which could be proportional to the renormalizable ${\cal
O}(1)$ spurions and powers of the cutoff, $\Lambda$---at every loop
order.\footnote{The non-renormalizable Yukawa interactions,
particularly the bottom interaction, will introduce quadratic
divergences. These contributions, however, are always small, as the
operators are suppressed by appropriate powers of the cutoff and $y_B
\sim 1$. We neglect their effect in the discussion that follows.} The largest corrections to the Higgs mass are from
logarithmically divergent operators, which always yield corrections to Higgs masses
that are proportional to $f_1^2/(4 \pi)^2 $ only.

We focus on operators generated in the full, unbroken $SU(4) \times
U(1)_X$ gauge theory from the renormalizable interactions. The operators generated in the unbroken $SU(4) \times U(1)_X$ break into four representative types described below. (The notation is  $M,N \in \left\{A,B,C,D\right\}$ and $\phi_M$ can refer to either $\phi_M$ or ${\phi'}_M$.)

\begin{itemize}
\item{ $\phi^{\dagger}_M \phi^{}_M$} 
\end{itemize}

These dimension-two operators (e.g. $\phi^{\dagger}_A \phi^{}_A$ or
${\phi'}^{\dagger}_A {\phi}^{}_A$) are already present in the theory
as mass terms for the $\phi^{}$ fields.  No masses for $H_1 \ \& \
H_2$ are generated by these terms, since both preserve the largest
global symmetry, $U(4)^4$, under which $H_1 \ \& \ H_2$ are
Goldstone bosons. Nevertheless, $\phi^{\dagger}_A \phi^{}_A$ can be
produced by quadratically divergent gauge, scalar, and fermion loops
which set the naturalness scale for the mass of the $\phi^{}$ fields
and the breaking scale $f_1$. The contributions are listed in Appendix
\ref{sec:qd}, and give \be m^2_{\phi} \simlt 10 \left(
\frac{\Lambda}{4\pi} \right)^2.  \ee The largest contribution comes
from the $\sqrt{2}$ enhanced top Yukawa ($y_T(1 \tev) \sim 1.25$),
while gauge and scalar loops contribute about a $1/3$ as much.
\begin{itemize}
\item{$\phi^{\dagger}_M \phi^{}_N$} 
\end{itemize}

These dimension-two operators (e.g. $\phi^{\dagger}_A \phi^{}_B$) will generate masses for $H_1 \ \& \
H_2$, since they break the $SU(4)^4$ global symmetries. However, $\phi^{\dagger}_A \phi^{}_B$  violates
the unbroken $U(1)_A$ and $U(1)_B$ symmetry, so it can only be
proportional to those spurions which also violate $U(1)_A$ and
$U(1)_B$, or is {\it the} spurion of this symmetry violation.  Setting $b_0$ at
$\sim .1 \tev$, these operators are safely ignored. Note that as long as all of
the dimensionless ${\cal O}(1)$ spurions preserve the $U(1)^4$ symmetry, 
they cannot generate $\sim \Lambda^2$ contributions to these operators.

We could also consider operators of the form $\left(\phi_M^{\dagger} \phi^{}_N\right)\left( \phi_P^{\dagger} \phi^{}_P \right)$ which also break the $U(1)$ global symmetries.  These
will be generated by finite loops involving $\phi^{\dagger}_M \phi^{}_N$ and are naturally
small.

\begin{itemize}
\item{$\left(\phi_M^{\dagger} \phi^{}_M\right)\left( \phi_N^{\dagger} \phi^{}_N \right)$}
\end{itemize}

These dimension-four operators (e.g. $\left(\phi_A^{\dagger} \phi^{}_A\right)\left( \phi_B^{\dagger} \phi^{}_B \right)$) preserve the global $U(4)^4$ symmetry, so do not lead to masses for $H_1 \ \& \ H_2$. 

\FIGURE[t]{ \epsfig{file=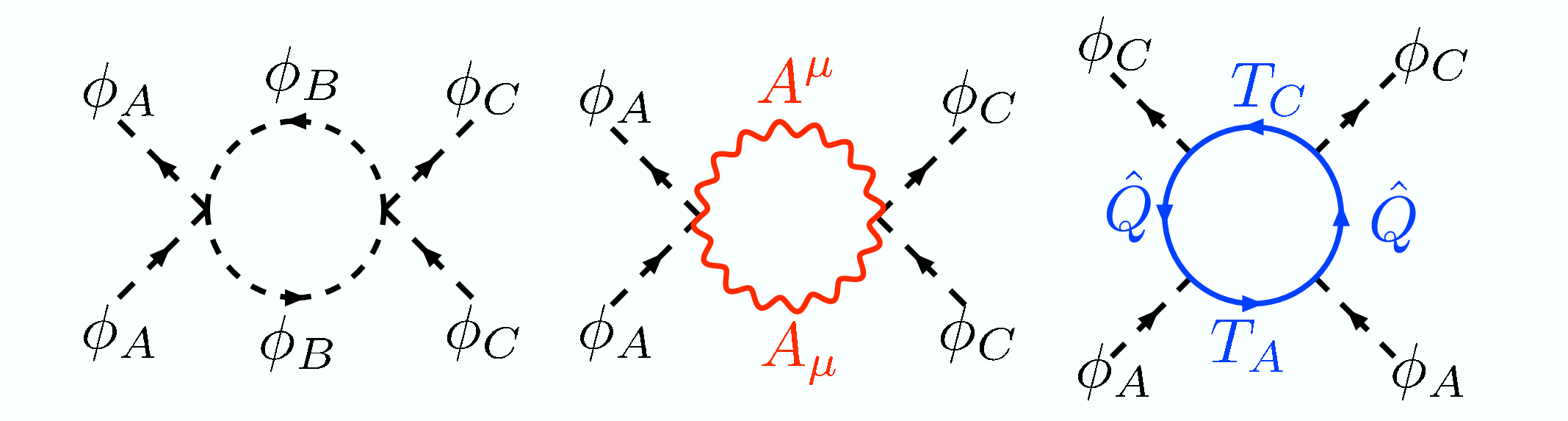,width=5in}
\caption{The three types of loops (scalar, gauge, and fermion loops,
respectively) that lead to logarithmically divergent dimension-four
operators. These operators contribute corrections to the masses of
the pseudo-Goldstone bosons of order $f_1^2/ 16 \pi^2$. }
\label{fig:d4}
}

\begin{itemize}
\item{$\left( \phi_M^{\dagger} \phi^{}_N \right)\left(\phi_N^{\dagger} \phi^{}_M \right)$}
\end{itemize}

 Consider the operator $\left( \phi_A^{\dagger} \phi^{}_C\right)
\left( \phi^{\dagger}_C \phi^{}_A \right)$.  This dimension-four
operator preserves the $U(1)_A \ \& \ U(1)_C$ symmetries, so it can be
generated by the ${\cal O}(1)$ spurions ($g, y_T, \lambda$). It also
violates the $SU(4)^4$ symmetry, so it is relevant for producing a
mass for $H_1$. However, contributions to this operator are only
logarithmically divergent (the coupling is dimensionless), and do not yield quadratic sensitivity to
$\Lambda$. The diagrams that generate these operators at one loop are
shown in Figure \ref{fig:d4}. These are the most divergent renormalizable
operators one can write down that preserve the $U(1)^4$ symmetries and
generate masses for the low-energy Higgs sector.

After expanding the fields in terms of their Higgs components, we find that the typical dimension-four operator predicts
\be
\frac{\kappa}{8 \pi^2} \left| \phi_A^{\dagger} \phi^{}_C \right|^2 \rightarrow m^2_{H_1} \sim \kappa \left(\frac{f_1}{4 \pi}\right)^2.
\ee
Operators with $A \rightarrow \{A, A'\}$ and $C \rightarrow \{C, C'\}$ (four operators in total) can also give contributions to $m^2_{H_1}$ of the same size, while operators with $A \rightarrow \{B, B'\}$ and $C \rightarrow \{D, D'\}$ contribute to $m^2_{H_2}$. 

We estimate the size of the loop contributions to these operators by
computing the logarithmically divergent piece of the effective
potential, with couplings constants evaluated at $1 \tev$ using the
RGE's given in Appendix \ref{sec:RGE}.  The largest contributions at
this stage are from scalar and fermion loops, while gauge loops are
smaller due to the small gauge couplings.

The number of scalars which contribute to the loops
shown in Figure \ref{fig:d4} is determined by the number of indirect
links between the $\phi^{}_A$ and $\phi^{}_C$. For this stage, there
are four such links (one each through $\phi^{}_B, \phi'_B, \phi^{}_D,\
\& \ \phi'_D$), which gives a factor of $4^1$, \be \kappa_s =
\left(\ln{10}\right)\ 4^1 \lambda^2 \rightarrow m^2_H \sim 9 \left( \frac{f_1}{4 \pi} \right)^2.
\ee
In the estimate for $m_H^2$ we have included the contribution from all four operators (those with $A \rightarrow A'$ and/or $C \rightarrow C'$) and used the running value of $\lambda(1 \tev) \sim .5$, assuming a low energy value $\lambda(.1 \tev) \sim .5$. It is important to note that the fine-tuning for this (and all) stages is associated with the Higgs mass squared $= \lambda v$ 
where $v$ is the electroweak vev.

Fermions give a contribution which scales with $N_c$, 
\be \kappa_f
\sim 2 N_c y_{TA}^2 y_{TC}^2 \ln{10}\rightarrow m^2_H \sim 33 \left(
\frac{f_1}{4 \pi} \right)^2.  \ee
Again, we have included the
contribution from all induced operators in the estimate for $m^2_H$ and used the running value of the top Yukawa $y_T( 1 \tev) \sim 1.25$. The top Yukawa is larger than 1 due to the $\sqrt{2}$ normalization factor described below Equation \ref{eq:tope}.  These corrections are larger than those present in the non-linear sigma models because of the enhancement of the top Yukawa and the presence of non pNGB scalar fields.  This is worse than the minimal linear sigma model
completion of the $SU(4)$ little Higgs theory because our completion has {\it two} scalars in the 
fundamental representation for each vev in the non-linear sigma model ({\it e.g.} fields $A$ and
$A'$).  The doubling is in anticipation of the tower:  the little Higgs stages produce two Higgs 
doublet models in the IR, so the linear sigma models which produce the symmetry breaking should also be ``two-scalar fundamental models'', in analogy with two-Higgs doublet models.

%%%%%%%%%%%%%%%%%%%%%%%%%%%%%%%%%%%%%%
%%%%%%%%%%%%%%%%%%%%%%%%%%%%%%%%%%%%%%
\section{Below 100 \tev}%%%%%%%%%%%%%%%%%%%%%%%%%%
\label{sec:100}%%%%%%%%%%%%%%%%%%%%%%%%%%%%%%
%%%%%%%%%%%%%%%%%%%%%%%%%%%%%%%%%%%%%%

\FIGURE[t]{\epsfig{file=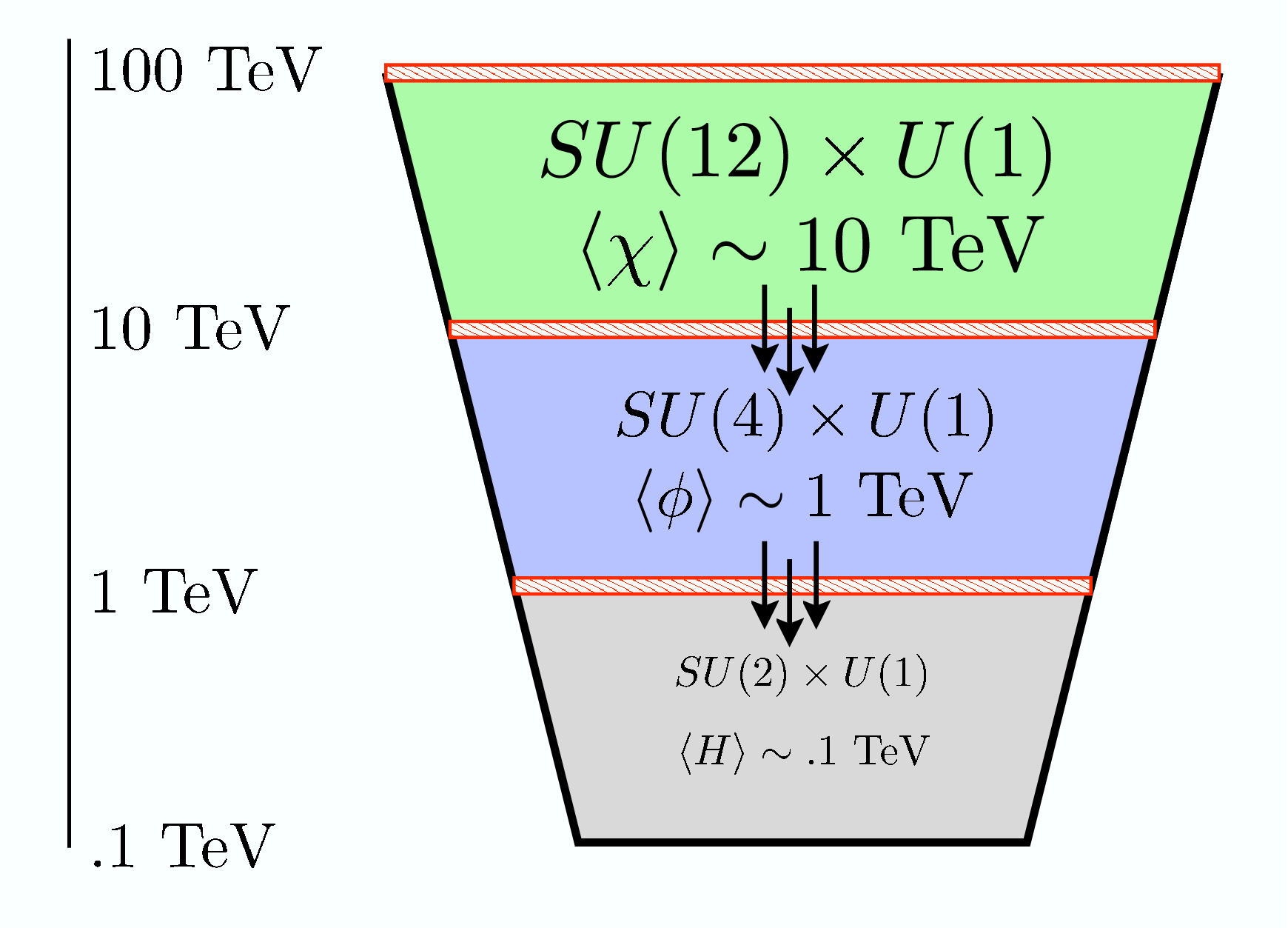, width=3.5in}\caption{The 
energy scale and structure of a two-stage Little Higgs theory. As in
Figure \ref{fig:10}, the theory can be represented by successive
stages of EFT's: A spontaneously broken $SU(12) \times U(1)$ gauge
theory lives between $10 \ \& \ 100 \tev$, leaving behind a
spontaneously broken $SU(4) \times U(1)$ gauge theory that lives
between $1 \ \& \ 10 \tev$, which leaves behind a spontaneously broken
$SU(2) \times U(1)$ gauge theory that lives between $.1 \ \& \ 1
\tev$.} \label{fig:100}}

As promised we now try to lift the cutoff to $100 \tev$, but we meet
with immediate obstacles. Absent new interactions, fine-tuning first
creeps in through the mass of the $\phi^{}$ fields; Unlike the
electroweak scale, the scale of $SU(4)$-breaking is sensitive to
quadratic divergences proportional to the cutoff which generate the
operators $\phi^{\dagger}_M \phi^{}_M$. Even though the electroweak
scale, $f_0$, remains insensitive to quadratic divergences, $f_0$
retains sensitivity to $f_1$. As $\Lambda$ is pushed to $100 \tev$, fine-tuning is introduced into the weak-scale.

Therefore, to push the cutoff, $\Lambda$, up to $100 \tev$ requires
introducing new interactions if the theory is to remain natural. These
new interactions should remove quadratic divergences to the $SU(4)$
breaking scale, $f_1$, without reintroducing unnatural scale
dependence into the electroweak breaking scale, $f_0$. 

We attempt to satisfy this stringent requirement by embedding the $SU(4) \times
U(1)_X$ gauge theory into an $SU(12) \times U(1)_Z$ gauge theory. This
gauge theory is cutoff by the scale $\Lambda \sim 100 \tev$, and
interactions are such that the $SU(12) \times U(1)_Z$ gauge theory is
collectively and spontaneously broken at the scale $f_2 \sim 10 \tev$
by a set of fundamental scalars, $\chi$. The $SU(4) \times U(1)_X$
gauge theory of Section \ref{sec:10} is left over, with the fields
$\phi^{}$ surviving as light pseudo-Goldstone bosons without quadratic
sensitivity to $\Lambda \sim 100 \tev$. All required $SU(4)$
quartics -- both those that directly stabilize $\phi$ vev's and those
that will eventually stabilize the Higgs vev -- are again generated at
tree-level. All other quartic interactions (which we assumed were absent in the $SU(4)$
theory) are generated at 1-loop order and are suppressed.  However, the suppression
is not quite large enough, due to the compensating factor of the number of fields, and
some fine-tuning remains at this stage.  We
review the fields and scales involved in Figure \ref{fig:100}.

%%%%%%%%%%%%%%%%%%%%%%%%%%%%%%%%%%%%%%
\subsection{$SU(12)\times U(1)_Z$ Breaking}%%%%%%%%%%%%%%%

\FIGURE{ \epsfig{file=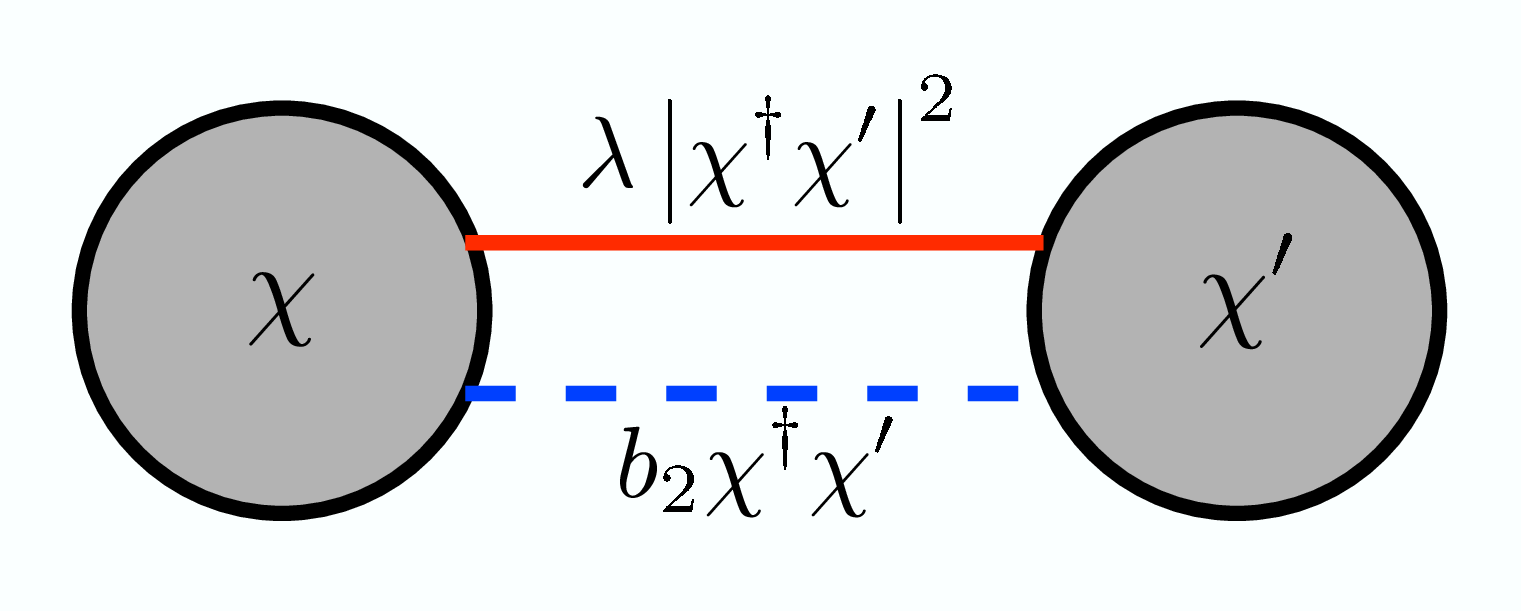,width=2in}
\caption{}
\label{fig:2chi}
} 

We build the $SU(12)\times U(1)_Z$ sector up iteratively from the
$SU(4)\times U(1)_X$ theory. The basis of this pattern is again the
unit shown in \ref{fig:2chi} which now visually depicts two
fundamental scalars $\chi \ \& \ \chi'$ of the $SU(12)$ gauge theory
and their interactions, and we have set $b_2 \sim 10 \tev$. We
proceed to build up a pattern of squares based on this unit, such that
once spontaneous symmetry breaking occurs in the $SU(12)$ theory, we
are left with the pattern of Figure \ref{fig:8} in the (low energy EFT) $SU(4)$ gauge theory below $10 \tev$. 

A single pair of $\chi$ fields produces a set of $U(12)/U(11)$
would-be Goldstone bosons along the flat directions of the $\chi$
potential. These would-be Goldstone bosons are actually eaten by the
gauge bosons, as the gauge symmetry is broken down to $SU(11) \times
U(1)$.

A single {\it square} of $\chi$ fields (four pairs, with scalar
interactions just like those of Figure \ref{fig:8} with $\phi
\rightarrow \chi$) leaves a pair of pNGB, $\phi$. The counting is
identical to that of Section \ref{sec:10}: each square of $\chi$
fields breaks the rank of the gauge group by 2; each square produces 2
would-be Goldstone fundamentals of $SU(4)$ that are eaten by the
$SU(12)$ gauge bosons; each square produces another 2 pNGB
fundamentals of $SU(4)$ that remain light and become a pair of $\phi$
fields in the $SU(4)$ EFT.

The full $SU(12)\times U(1)_Z$ gauge theory contains 4 such squares,
to generate the 8 light pNGB scalars, $\phi$ in the $SU(4) \times
U(1)_X$ theory. We require additional quartic interactions between
{\it squares} as indicated in Figure \ref{fig:32to8}. These quartic
links make it more energetically favorable for the vev's in each
square to separate and break the gauge symmetry to $SU(4)$, not
$SU(10)$. The net result is that the fields $\chi$ occur in sets of
four (these are the two pairs on opposite corners of a smaller square)
with vev's in the same position. Each set of four has a vev in a
unique position, which breaks the gauge group from $SU(12) \times
U(1)_Z$ to $SU(4) \times U(1)_X$ and produces the 8 pNGB, $\phi$. The
alignment of vev's within the small squares is guaranteed as in
Section \ref{sec:10}, by the positive quartics and the introduction 
of small $b$ terms and quartics across the small square. The (mis)
alignment of vev's between fields across the largest square can be
guaranteed by the outer positive quartics and the even smaller $b$ terms
and quartics introduced across the largest square to ensure subsequent alignment in
the $SU(4)$ theory.  

Together, the quartic links collapse to produce the quartic
interactions as shown in Figure \ref{fig:32to8}. As the vertices of
each small square collapse to become a pair of $\phi^{}$ fields, the
edges collapse to become the quartic interaction for the pair. The
largest square of links collapses to become the square of links in the
$SU(4)$ gauge theory that leads to $SU(4)$ quartics and the final
quartic in the $SU(2)_W$ 2HDM. The mechanism that accomplishes this is
identical to that of Section \ref{sec:10}--and occurs at
tree-level. The low-energy quartics all have the same value as the
original $SU(12) \times U(1)_X$ quartics and come from integrating 
out scalars with trilinear interactions and masses of order $f_2$.

\FIGURE[t]{\epsfig{file=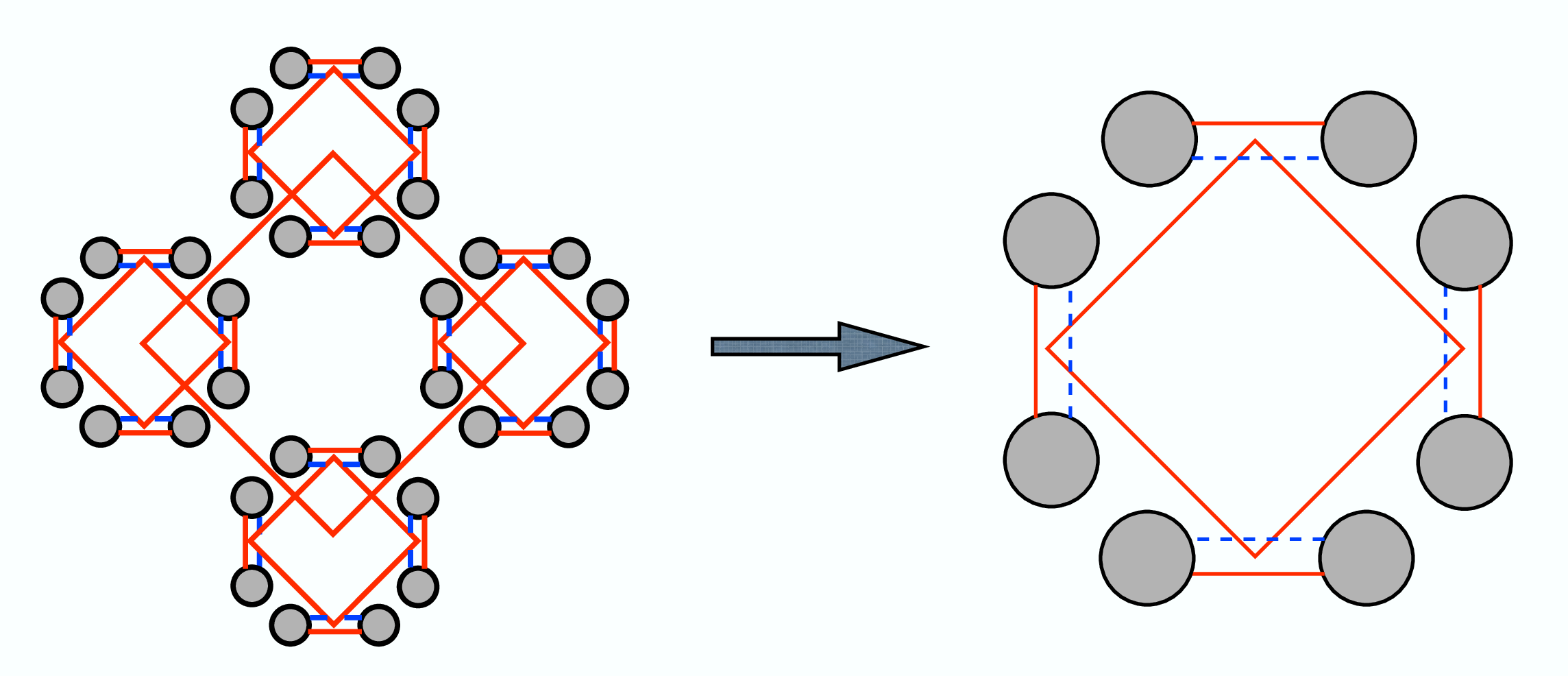, width=5in}\caption{The $SU(12)$
gauge theory (left) juxtaposed with the $SU(4)$ gauge theory
(right). Every pair of fields in the $SU(4)$ theory is replaced with a
square of 8 fields in the $SU(12)$ theory. Shaded circles are again 
fields that get vev's, while red, solid lines represent $U(1)$ preserving 
quartic interactions between fields at opposite ends of the line, and the 
blue dashed lines represent $U(1)$ breaking dimension-two operators 
($b$-terms) between fields at opposite ends of the line.} \label{fig:32to8}}

The $b_1$ and $b_0$-terms that promote $SU(4)$ and $SU(2)_W$ breaking,
respectively, are added in by hand by linking the relevant $\chi$
fields. Such links are again technically natural in the larger theory. All 
singlets that are would-be Goldstones of the spontaneously broken 
$U(1)_{\chi}$ global symmetries are eaten by $SU(12) \times U(1)_Z$ 
gauge bosons or pick up $\sim f_1$ masses from the explicit symmetry 
breaking terms $b_1$.

The theory begins with a large $SU(12) \times U(1)^{16}$ global
symmetry that is preserved by gauge, scalar, and Yukawa interactions.
When the $\chi$ fields acquire vev's, this symmetry is broken to
$SU(4) \times U(1)^8$, which is further broken to $SU(4) \times
U(1)^4$ by the $b_1$ interactions. This is exactly the global symmetry
we used to show the absence of all dangerous dimension two operators
in the $\phi$ theory below $10 \tev$.  

%%%%%%%%%%%%%%%%%%%%%%%%%%%%%%%%%%%%%%
\subsection{$U(1)_Z$ \& Fermions} %%%%%%%%%%%%%%%%%%%%
\label{sec:u1z} %%%%%%%%%%%%%%%%%%%%%%%%%%%%%%

All scalars $\chi$ have charge $Z_{\chi}=-1/12$ under $U(1)_Z$, and the
$U(1)_Z$ mixes with one of the diagonal generators of $SU(12)$ to produce
$U(1)_Z$.

Specifically:
\be
U(1)_X = -\frac{1}{12}{\rm diag} \{2,2,2,2,-1,-1,-1,-1,-1,-1,-1,-1\} + U(1)_Z
\ee
for a fundamental of $SU(12)$.  We note that 
\be
g_X=\frac{g g_Z}{\sqrt{g^2 + \frac{g_Z^2}{3}}}.
\ee

We embed fermions as described in Section \ref{sec:10}: the third
generation has interactions that remove quadratic divergences from the
top sector only (with third generation fermion charges $W_{{\tilde{Q}}_3}=7/12$), while the first two
generations remove divergences from the down-type quarks and
neutrinos ($W_{{\tilde{Q}}_{1,2}}=-3/12$, $W_{\tilde{L}}=-1/12$). 

In Section \ref{sec:u1x}, we used a set of Yukawa interactions
that caused the Yukawa couplings to be enhanced by a factor of
$\sqrt{2}$ in the $SU(4) \times U(1)_X$ stage. To keep the Yukawa
interactions under control in the $SU(12)$ theory, we write down an
interaction of the form $\left(y \chi+ {y'} {\chi'} \right)\tilde{T}
\tilde{Q}$ instead of the single interaction $y \chi \tilde{T}
\tilde{Q}$. This embedding causes a reduction in the value of the
$SU(12)$ Yukawa by a factor of $\sqrt{2}$, and brings the top Yukawa
under perturbative control.  However, the $U(1)_{\chi}$ and
$U(1)_{\chi'}$ symmetries which are only broken by $b_2 \sim 10 \tev$
terms are now also broken by dimensionless combinations of $y$ and
$y'$. This has the advantage of generating the $b_2$ term naturally,
but makes it difficult to write down simple UV completions of the
fermion sector that will not introduce dangerous scale dependence. 
We therefore only use this embedding in the $SU(12)$ stage.

This embedding is again anomaly free. The non-renormalizable
interactions needed for the bottom quark now involve 11 $\chi$ fields
and have overall coefficient, $y_B/ f_2^{10}$. Since the mass
suppression is $f_2$ and not $\Lambda$, the bottom Yukawa forces us to
a strong-coupling limit at the cutoff that we address in Section
\ref{sec:nl}. To add another stage would require a new embedding for
both the top and bottom quarks. 

%%%%%%%%%%%%%%%%%%%%%%%%%%%%%%%%%%%%%%
\subsection{Loop Corrections} %%%%%%%%%%%%%%%%%%%%%%%
\label{sec:100qc} %%%%%%%%%%%%%%%%%%%%%%%%%%%%%

After the $SU(12)$ symmetry breaking, the $\phi$ fields remain
massless at tree-level (up to small $b$ terms), but we must again consider masses that might
be generated at loop level. The global $U(1)_{\chi}$ symmetries again
prevent the most dangerous dimension two-operators which could give
$\sim \Lambda^2$ masses to the $\phi$ fields from the renormalizable
${\cal O}(1)$ spurions at all loop orders. Here we focus on the form of
corrections from just the renormalizable sector.

The largest corrections to the $\phi$ masses come from
logarithmically divergent operators that give masses proportional only
to $f_2^2$. The $U(1)$ symmetry structure also prevents dangerous
contributions to the {\it Higgs} masses: No masses are generated which
are proportional to the ${\cal O}(1)$ spurions times powers of
$\Lambda^2$ or $f_2^2$, again at all loop orders.

The types of operators that we've ignored in the full $SU(12) \times U(1)_X$ 
theory are ($J,K$ run over all 32 $\chi$ fields)

\begin{itemize}
\item{ $\chi^{\dagger}_J \chi^{}_J$ }
\end{itemize}

These dimension-two operators are already present in the theory as
mass terms for the $\chi^{}$ fields. No masses for $\phi$ are
generated by these terms, since $\chi_J^{\dagger} \chi^{}_J$ preserves
the largest global symmetry, $U(12)^{16}$, under which the $\phi$ fields
(and therefore, the Higgs fields) are exact Goldstone
bosons. Nevertheless, these operators are generated by quadratically
divergent gauge, scalar, and fermion loops and set the naturalness
scale for the mass of the $\chi^{}$ fields and the breaking scale
$f_2$. The contributions are listed in Appendix \ref{sec:qd}, and give

 \be m^2_{\chi^{}} \simlt  7 \left( \frac{\Lambda}{4 \pi} \right)^2,  \ee
The contribution grows with the number of scalar fields and the size 
of the weak gauge group, but is offset by the running coupling 
constants ($g_w(10 \tev) \sim .6, \lambda(10 \tev) \sim .3$), now assuming a low
energy value of $\lambda\sim .7$.
The fermion contribution is reduced compared to that in the 
$SU(4) \times U(1)_X$ theory because of the {\it reduction} in the 
Yukawa couplings from the doubling of interactions, so 
$y_T(10 \tev) \sim .9$. In this stage, the $b_1$ terms are also 
generated with the same size by the enhanced fermion structure 
that mixes the primed and unprimed fields.

\begin{itemize}
\item{$\chi^{\dagger}_J \chi^{}_K$} 
\end{itemize}

These dimension-two operators will generate masses for both the $\phi$
sector and the Higgs sector, since they break the $SU(12)^{12}$ global
symmetries.  However, these operators violate the unbroken $U(1)^{16}$
symmetries, so they must be proportional to those spurions which also
violate the relevant $U(1)$ symmetries. The only coefficients which
violate these symmetries are the $b_1$ terms (which can only generate
masses for the $\phi$ fields proportional to $b_1$), and the $b_0$
terms (which can generate masses for the $\phi$ fields and the Higgs
fields). Both sets of $b$ terms give corrections that are under control.
As in the $SU(4)$ stage, operators of the form
$\left(\chi_J^{\dagger} \chi^{}_K\right)\left( \chi_L^{\dagger} \chi^{}_L \right)$
will be generate by finite loops and will be small ({\it i.e.}, have a negligible 
effect on the $\phi$ masses).

\begin{itemize}
\item{$\left(\chi_J^{\dagger} \chi^{}_J\right)\left( \chi_K^{\dagger}
\chi^{}_K \right)$} 
\end{itemize}

These dimension-four operators preserve the global
$U(12)^{16}$ symmetry under which the $\phi$ (and Higgs) fields are 
exact Goldstone bosons, so no mass terms are generated by these operators.

\begin{itemize}
\item{$\left( \chi_J^{\dagger} \chi^{}_K \right)
\left(\chi_K^{\dagger} \chi^{}_J \right)$} 
\end{itemize}

These dimension-four operators preserve the $U(1)^{16}$ global
symmetries, so they are generated by the ${\cal O}(1)$ spurions ($g,
y_T, \lambda$). The operators also violate the $U(12)^{16}$
symmetries, so they are relevant for producing a mass for $\phi$
proportional to $f_2^2$ after the vev's for $\chi$ are plugged in.
Operators which gives $f_2^2$ masses to the $\phi$ fields are quartic
interactions between $\chi$ fields which have vev's in the same
position.

These operators are only logarithmically divergent, and do not yield 
quadratic sensitivity to $\Lambda$. The diagrams that generate these
operators are identical to those shown in Figure \ref{fig:d4}, except now 
involving $\chi$ fields and the $SU(12)$ gauge bosons and quark fields. 
These are the only renormalizable operators one can write down that 
preserve the $U(1)^{16}$ symmetries and generate masses for the low-energy 
$\phi$ sector.

None of these operators, however, generate $f_2^2$ masses for the
Higgs sector. This can be seen in two ways: if one
carefully expands the $\chi$ fields in terms of the Higgs fields, it
can be seen that the $\chi$ vev's never generate a Higgs mass term; or
if one notes that the EFT below the $\chi$ vev's contains an unbroken
$U(1)^4$ symmetry on the $\phi$ fields that prevents all dimension two
operators that lead to Higgs masses proportional to $f_2$. These are
the same $U(1)$ symmetries we used to explain the absence of large
$\phi^{\dagger}_M \phi_N^{}$ terms in Section \ref{sec:10qc}.

After expanding the fields in terms of their Higgs components, we find
that the typical dimension-four operator predicts \be \frac{\kappa}{8
\pi^2} \left| \chi_J^{\dagger} \chi^{}_K \right|^2 \rightarrow
m^2_{\phi} \sim \kappa \left(\frac{f_2}{4 \pi}\right)^2.  \ee 
For each field $\phi_M$, there can be four operators which contribute to its mass. 

The number of scalars which contribute to the loops shown in Figure
\ref{fig:d4} is determined this time by the number of indirect links
between the $\chi$ fields. For this stage, there are 20 such links
(four from within the little square, and 8 each from the neighboring
squares): \be \kappa_s = \left( \ln{10} \right)\ \left( 4^1 + 4^2
\right) \lambda^2 \rightarrow m^2_{\phi} \sim 15 \left( \frac{f_2}{4
\pi} \right)^2.  \ee In the estimate for $m_{\phi}^2$ we have the
contribution from all four operators and used the running value of
$\lambda(10 \tev) \sim .3$, assuming a low energy value $\lambda(.1
\tev) \sim .7$. The quartic coupling is running weak due to
contributions from the top and gauge sectors. Gauge loops
contribute corrections that are a bit smaller in size, while fermion
contributions are larger \be \kappa_f
\sim 2 N_c y^4 \ln{10}\rightarrow m^2_{\phi} \sim 36 \left(
\frac{f_2}{4 \pi} \right)^2.  \ee 
Again, we have included the contribution from the four separate
operators that can contribute in the estimate for $m^2_{\phi}$, and
used the running $y_T(10 \tev) \sim .9$.  We have not included other 
potential contributions from the other couplings added previously
to reduce the couplings from $\sim\sqrt{2}$ to $\sim 1$, and thus this
estimate may be low.  While all of these
logarithmically divergent terms give the right parametric dependence,
fine-tuning is introduced by typically neglected numerical prefactors: 
the number of scalar fields, rank of the gauge group, and combinations 
of $N_c$ and the large Yukawa couplings. 

The operators just described lead to masses for the $\phi$ fields of
order $f_2$, but there are also operators (produced in the $SU(12)$
theory) that can contribute to $Higgs$ masses of order $f_1$ when the
$SU(4)$ gauge theory breaks. Such operators are quartics that mix a
field from one square, from any other field on the diametrically
opposed square. These operators do not give $\sim f_2$ masses to the
Higgs since the vev's of these two fields miss (because of the
preserved $U(1)_{\phi}$ symmetries), but they do lead to operators
like $\left| \phi^{\dagger}_A \phi^{}_C \right|^2$ that we considered
at the end of Section \ref{sec:10qc}.

We assumed these operators were absent at tree-level in the $SU(4)$
theory of Section \ref{sec:10} and must show that the $SU(12)$ UV
completion does not generate them with large coefficients. These
dimension-4 operators are generated at 1-loop order by logarithmically
divergent diagrams. The largest of these comes from Yukawa
interactions, and the coefficients generated have size $\sim .1$,
while contributions from scalar operators are of order $.04$.  These
quartics are $10-30 \%$ perturbations compared to the tree level
quartics and are of the same size as the corrections calculated at the
end of Section \ref{sec:10qc}.

%%%%%%%%%%%%%%%%%%%%%%%%%%%%%%%%%%%%%%
\subsection{A Non-Linear Sigma Model} %%%%%%%%%%%%%%%%%%
\label{sec:nl} %%%%%%%%%%%%%%%%%%%%%%%%%%%%%%%

The non-renormalizable bottom Yukawa (requiring an epsilon tensor and
11 $\chi$ fields) is a dimension-14 operator which would produce way too
small of a bottom quark mass unless the theory was strongly coupled at $\Lambda$.
Thus, we are forced into the non-linear sigma model
where the radial modes including in the $\chi$ fields are pushed to
the cutoff. In this case, we keep the same interactions as before,
except now each $\chi$ field is a dimensionless non-linear sigma model
field (see \cite{Kaplan:2003uc} for the analogous discussion in the $SU(4)\times U(1)$ model). The coefficient of
the non-renormalizable bottom interaction is $\sim 1$ times the appropriate powers
of $f_2$ in order to produce the correct standard model Yukawa coupling.

The non-linear sigma model has the significant advantage of removing $1/2$ 
of the scalar fields since only one $\chi$ field is needed to describe each 
pNGB $\phi$ field (instead of the two fields $\chi, \chi'$. This reduces the 
effect of the logarithmic divergences by 1/2, so the NDA estimate for the 
fine-tuning is similarly reduced.

%%%%%%%%%%%%%%%%%%%%%%%%%%%%%%%%%%%%%%
%%%%%%%%%%%%%%%%%%%%%%%%%%%%%%%%%%%%%%
\section{Dreams of a Goldstone Tower} %%%%%%%%%%%%%%%%%%
\label{sec:stack} %%%%%%%%%%%%%%%%%%%%%%%%%%%%%
%%%%%%%%%%%%%%%%%%%%%%%%%%%%%%%%%%%%%%

Up to this point we have described a two-stage Little Higgs model with a cutoff of $100 \tev$.  
We have successfully removed all dangerous loop contributions -- 
there are no quadratic divergent contributions to the Higgs mass and all 
log-divergent and finite contributions are proportional to only the lowest breaking scale $f_1$.
At the same time, large numerical factors render the ``safe'' loop contributions
unpalatable by reintroducing fine tuning.  The existence of these large numbers
is due in part to the large coupling times color factor in the top sector and due in part
to the geometric increase in the number of fields.  In spite of this drawback, the natural
repeatability of the structure of our theory compels us to explore the possibility of a
full tower of symmetry breaking.  We find a toy model of a tower with reduced fine-tuning
and indications of how one might remove the tuning completely.

The stacking of little Higgs modules is quite easy.  Every 
two-scalar-fundamental sector of the theory is completed by eight fundamentals 
of the larger gauge group, as the original two-Higgs-doublet model of 
$SU(2)\times U(1)$ completes into an $SU(4)\times U(1)$ gauge theory with 
eight fundamentals.  At each stage the number of scalars in the fundamental 
representation increases by a factor of four, {\it i.e.},
\begin{equation}
N_s = 2 \times 4^k,
\end{equation}
where $k$ labels the stage of breaking, $f_k$:  $f_0=100 \ {\rm GeV}$, 
$f_1= 1 \tev$, $f_k = 10^{k-1} \tev$.  Each one of the groups of eight 
fundamentals which produces a two-scalar-fundamental model breaks 
the rank of the gauge group by 2.  Thus, the rank of the group in the 
upper stage is equal to the rank of the group in the stage just below 
it plus the number of fundamental scalars.  For example, $SU(2)$ with 
2 little Higgses required an $SU(2+2)$ theory with 8 fundamentals above 
it, which required an $SU(4+8)$ theory with 32 fundamentals above that, etc. 
(see Table \ref{table:stack}).  This counting also makes it clear that at every 
stage, there are exact (up to $b$ terms) $U(1)_{scalar}$ symmetries which 
rotate by a phase each of the $SU(N)$ fundamentals.  Each $U(1)_{scalar}$ at one 
stage is a linear combination of a broken diagonal generator of the gauge 
group in the upper stage and the $U(1)_{scalar}$ symmetry that rotates the 
phase of the linear combination of scalars responsible for the breaking in the upper stage.

\TABULAR[h]{|l|c|c|c|}{\hline Breaking Scale&$N_W$ & $N_S$& cutoff \\ \hline $f_0 \sim .1 \tev$ & 2 & 2 & $1 \tev$\\\hline $f_1 \sim \tev$& 4 & 8 & $10 \tev$ \\ \hline $f_2 \sim 10 \tev$ & 12 & 32 & $100 \tev$ \\ \hline $f_k \sim 10^{k-1} \tev $ & $2+\frac{2}{3}\left(4^k -1 \right)$ & $ 2 \cdot 4^k$ & $10^k \tev$\\ \hline}{}
\label{table:stack}

What is clear from this discussion (and the table) is that this theory quickly becomes a large $N$ theory (with $N_s \sim N$).  This could in principle be okay if everything has the right $N$-scaling behavior to keep the theory perturbative and functional.  In practice, this means gauge couplings squared $g^2$, Yukawa couplings squared $y^2$, and quartic couplings $\lambda$ must all scale like $1/N$.  Then one should analyzing the running and matching of the 't Hooft coupling, {\it e.g.}, $g^2 N$, to see if it remains perturbative in the UV.  Amazingly enough, in the simplest version of the field content (only $SU(N)$ gauge bosons and scalars), the average 't Hooft coupling approaches an asymptotic value that is perturbative!  See Figure \ref{fig:thooft}.
\FIGURE[t]{ \epsfig{file=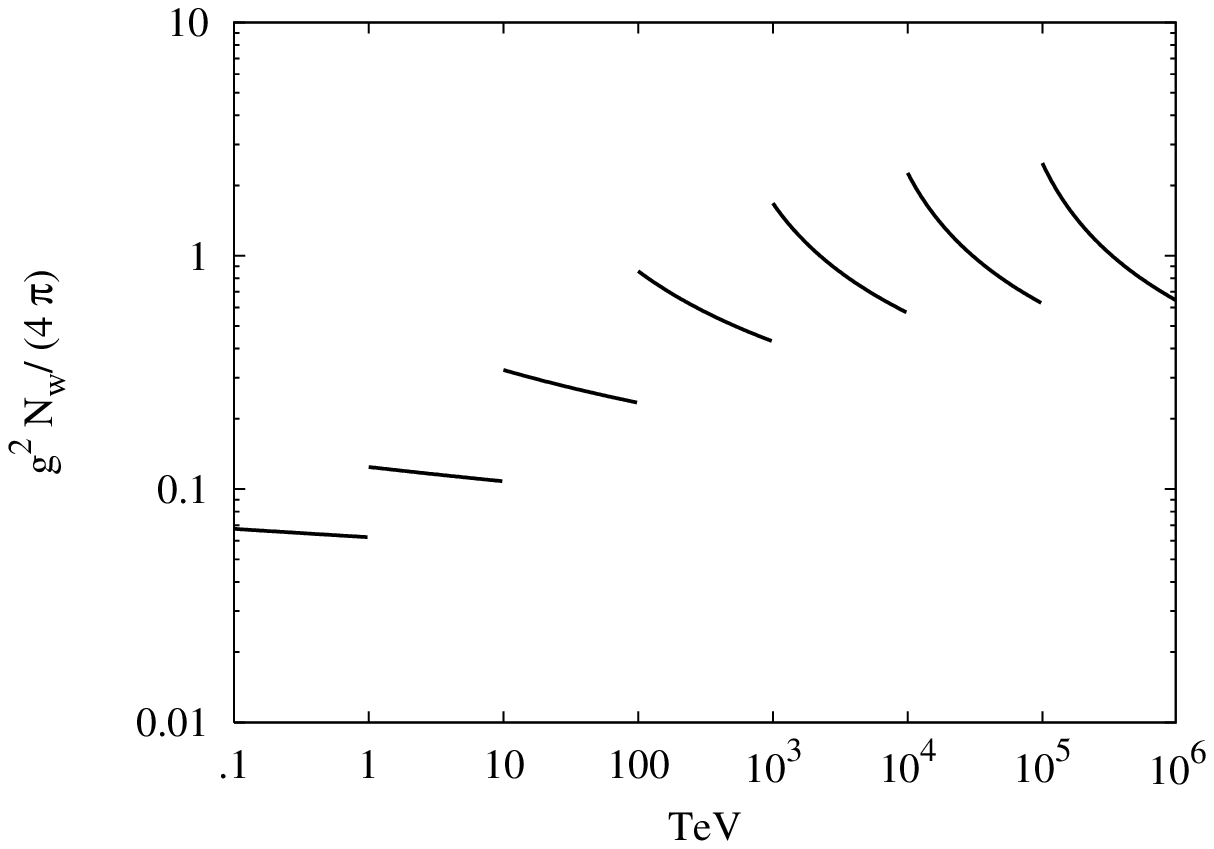,width=4in}
\caption{Evolution of the 't Hooft coupling, $g^2 N_w / (4 \pi)$ from the one renormalization group equations. The discontinuities are caused by the changing value of $N_w$ at every decade of energy.}
\label{fig:thooft}
}

However, there are some operators/couplings which do not obey the necessary 
scaling to preserve the structure.  The most obvious ones are the hypercharge 
and $SU(3)$ color couplings which, while obtaining beta functions which scale 
linearly with $N$, are asymptotically {\it unfree} and therefore hit a Landau pole 
at some intermediate scale (though they remain perturbative at 100 TeV).  Some 
level of unification of gauge groups would be required to avoid these problem in 
a model like ours.  A more immediate problem is the bottom Yukawa coupling 
(among others) which already become impossible to write down if the $SU(12)$ 
model below 100 TeV is a {\it linear} sigma model (what would be required to 
continue the tower).  In the linear sigma model, the bottom Yukawa requires an 
operator of dimension 14!  However, in order for Yukawas to follow true $1/N$ scaling, some 
fermions must appear in representations with two indices - {\it i.e.}, those which have
a Dynkin index which goes like $N$.  Unfortunately, this doesn't change the situation for
fermions like the bottom due to the requirement of the epsilon tensor and thus remains a 
stopping point for realistic towers in this class of models. 

The above issue we regard as peripheral to the central point of discovering a perturbative theory with order one couplings in the infrared in which naturally light scalars appear. If we shed things like hypercharge and Yukawa couplings and focus on the structure of the scalar-gauge theory we find indications that a full tower may be possible.  For example, at every stage there is a cutoff $\Lambda$ and a mass term for the scalar fields $m^2 |\phi|^2$ whose value is generated by a one loop quadratic divergence $m^2 \sim g^2 N \Lambda^2 / 16\pi^2$.  If $g^2 N \sim 1$ then our one loop suppression remains and $m\sim \Lambda/4\pi$.  The vev of $\phi$, $f_k$,  however should be (without fine tuning) $f_k \sim \sqrt{m^2 / \lambda} \sim \sqrt{N}\Lambda/4\pi$!  One might worry that such a vev would be out of the range of validity.  One hint that it may not be is that because all couplings of $\phi$ are of order $1/\sqrt{N}$, then this vev only generates masses for other fields ({\it e.g.}, gauge bosons) below the cutoff.  The renormalizable operators exhibit an approximate shift symmetry for $\phi$ only broken by effects suppressed by powers of $N$.  Treating these as spurions, the higher dimensional operators contributing to the potential would be functions of $\phi/(\sqrt{N}\Lambda)$ times loop factors at weak coupling.  Under these conditions, the validity of the effective theory is maintained.

The real sticking point comes from the renormalization group flow of the 
quartic operators detailed in Appendix \ref{sec:RGE}.  While there is a 
range of quartic and $g^2$ values for which the quartic's beta function 
is negative, it is not a stable range - in other words, the couplings flow 
away from these values at eventually the quartics become IR free and 
stop tracking the $g^2$ coupling.  This appears to be a problem with 
the actual coefficients in the beta functions and thus the stack may still 
be possible with some other quartic structure or gauge structure involving 
different representations.  This is a possibility we will continue to pursue.
We are not, however, far off.  As it stands, the quartics in our model track 
effectively for a cutoff $\Lambda \simlt 10^5 \tev$.  In addition, the so-called
``bad'' quartics at each stage run in such a way as to stay a loop factor below
the ``good'' quartics, as long as they start that way at the high scale.  This is
the sign that in the large-$N$ theory, large logs do not become an issue.

The other sticking point is the fact that high scales, $g^2 N/4\pi \equiv \alpha_{\rm 't Hooft}$ is not $\sim 1/4\pi$
but $\sim 1$.  While this is still semi-perturbative, it would naturally allow (without
any fine tuning) only a
factor of $\sqrt{4\pi}$ between scales (and not $4\pi$).  If we adjusted the ratio
$f_k/f_{k-1}$ to fit the asymptotic average 't Hooft coupling, the asymptotic value
becomes $\alpha_{\rm 't Hooft} \sim 4\pi/3$ where the perturbative expansion is
suspect.  This again simply becomes a question of more favorable beta functions.

Eventually, one would like to explain the structure of the $b$ terms, the explicit 
$U(1)_{scalar}$-symmetry
breaking occurring at every stage.  While it is technically natural to include these 
operators, it amounts to a very large number of ``$\mu$-term'' problems in analogy
with the minimal supersymmetric standard model.  This we see as the most difficult
task as we feel it should be dimensionful operators at each stage that performs the
role of explicit symmetry breaking.  However, we have not been able to rule out a
fix for this displeasing feature.

The final question to address is how to attach such a tower in the UV, which at this point is a toy model,
to a realistic theory in the IR.  Hypercharge and color would have to be included in the gauge
group ramp up -- {\it i.e.}, some sort of unified group which does not generate proton decay such 
as a Pati-Salam \cite{Pati:1974yy} or trinification \cite{trin} seed.  
The other difficulty is connecting to fermions.  Here
we could imagine stealing from extensions of technicolor theories and attempt to implement
something like extended technicolor \cite{Dimopoulos:1979es,Eichten:1979ah}
in which part of the gauge group runs a bit stronger than QCD
somewhat above the weak scale.  While not beautiful, this could be our first stab at an existence proof.

In the end we have shown that the $SU(4)$ little Higgs theory, in its weakly coupled linear-sigma model form, is a repeatable structure.  The field responsible for the higher $SU(4)$ breaking is in the same representation (fundamental) as the ``little'' field responsible for the $SU(2)$ breaking.  The little Higgs structure in this model comes from vev's in multiple individual representations and therefore it is possible to preserve many of these $U(1)_{scalar}$ global symmetries at each stage.  While a pure stack of our model with only gauge fields and scalars breaks down at a few stages, we have found no reason in principle why an arbitrarily large stack of little Higgs modules wouldn't work.  We feel it is of significant interest to find if such a field theory is possible as it should have a unique and interesting structure.  One could then ask if such a structure could naturally fall out of a real potential UV completion, {\it e.g.}, string theory.

\acknowledgments{We would like to thank R. Sundrum, T. Tait \& C. Wagner 
for useful comments, and J. Wacker for his visualization techniques.  PB, and 
work at ANL, is supported in part by the US DOE, Div.\  of HEP, Contract 
W-31-109-ENG-38.  DK is supported in part by the NSF and by the 
Department of Energy's Outstanding Junior Investigator program.}

%%%%%%%%%%%%%%%%%%%%%%%%%%%%%%%%%%%%%%
%%%%%%%%%%%%%%%%%%%%%%%%%%%%%%%%%%%%%%
\begin{appendix} %%%%%%%%%%%%%%%%%%%%%%%%%%%%%
%%%%%%%%%%%%%%%%%%%%%%%%%%%%%%%%%%%%%%

%%%%%%%%%%%%%%%%%%%%%%%%%%%%%%%%%%%%%%
\section{Scalar fields \& $SU(4) \times U(1)_X$} %%%%%%%%%%%%%%
\label{app:su4} %%%%%%%%%%%%%%%%%%%%%%%%%%%%%%

In this Appendix we describe the full pattern of symmetry breaking in
the $SU(4) \times U(1)_X$ theory.  Given the full potential of Section
\ref{sec:10}, with the 8 scalars, $\phi_A, \phi'_A, \phi_B, \phi'_B,
\phi_C, \phi'_C, \phi_D, \phi'_D$, we can rewrite these fields in
terms of a normalized mass-basis after symmetry breaking. The full
potential we consider is that of Figure \ref{fig:8}. 

The quartics between pairs make it more
energetically favorable for neighboring pairs to acquire vev's in
different positions. We can, without loss of generality, choose these
vev's in the third and fourth position which break the gauged symmetry
down to $SU(2)_W$. Further, since the vev's are in different positions, many of
the scalars contained in the four $\phi^{}$ fields remain
massless. Recall that with all interactions (gauge and quartic) turned
off, we expect $4\times 7 = 28$ Goldstone bosons (from the four sets
of $U(4)/U(3)$ directions in field space). With the interactions
turned on, 12 scalars are would-be Goldstone bosons (including two
$SU(2)_W$ doublets) that are eaten by the $\left(SU(4) \times
U(1)_X\right)/\left(SU(2)_W \times U(1)_Y\right)$ gauge bosons, 6
scalars (three complex singlets) pick up masses of order $f$ from the
quartic potential, and 10 pNGB ( $H_1,\ H_2$ and two real singlets)
receive no tree-level masses at all.

The two extra singlets are the non-eaten evidence of the 4
spontaneously broken global $U(1)$ symmetries. While the gauge
symmetry breaks from $SU(4) \times U(1)_X$ to $SU(2)_W \times U(1)_Y$,
the global symmetry breaks from $SU(4) \times U(1)^4$ to $SU(2) \times
U(1)^2$.  These singlets actually pick up weak-scale masses from the
$SU(2)_W\times U(1)_Y$ $b_0$ term described in Section \ref{sec:su4}.

\bea
\phi_A&=&\left( \begin{array}{c} \frac{1}{2} \left( \sqrt{2} H_{A} + {\bf H_1} +\pi_{1}\right)\\ f_1 + \frac{1}{2} \left( \rho_{A1r}+I \rho_{A1i}+ \rho_{A2r}+I\rho_{A2i} \right)\\ \frac{1}{2\sqrt{2}}\left( \sigma_E+\sigma_1+\sigma_2 +\sigma_3 +2\sigma_{A}\right) \end{array}\right), \\
\phi'_A&=&\left( \begin{array}{c} \frac{1}{2} \left( -\sqrt{2} H_{A} + {\bf H_1} +\pi_{1}\right)\\ f_1 + \frac{1}{2} \left( -\rho_{A1r}-I \rho_{A1i}+ \rho_{A2r}+I\rho_{A2i} \right)\\ \frac{1}{2\sqrt{2}}\left( \sigma_E+\sigma_1+\sigma_2 +\sigma_3 -2\sigma_{A}\right) \end{array}\right), \\
\phi_B&=&\left( \begin{array}{c} \frac{1}{2} \left( \sqrt{2} H_{B} + {\bf H_2} +\pi_{2}\right)\\ \frac{1}{2\sqrt{2}}\left( -\sigma^{\star}_E+\sigma^{\star}_1-\sigma^{\star}_2 +\sigma^{\star}_3 +2\sigma^{\star}_{B}\right) \\ f_1 + \frac{1}{2} \left( \rho_{B1r}+I \rho_{B1i}+ \rho_{B2r}+I\rho_{B2i} \right)\end{array}\right), \\
\phi'_B&=&\left( \begin{array}{c} \frac{1}{2} \left( -\sqrt{2} H_{B} + {\bf H_2} +\pi_{2}\right)\\ \frac{1}{2\sqrt{2}}\left( -\sigma^{\star}_E+\sigma^{\star}_1-\sigma^{\star}_2 +\sigma^{\star}_3 -2\sigma^{\star}_{B}\right) \\ f_1 + \frac{1}{2} \left( -\rho_{B1r}-I \rho_{B1i}+ \rho_{B2r}+I\rho_{B2i} \right)\end{array}\right), \\
\phi_C&=&\left( \begin{array}{c} \frac{1}{2} \left( \sqrt{2} H_{C} - {\bf H_1} +\pi_{1}\right)\\ f_1 + \frac{1}{2} \left( \rho_{C1r}+I \rho_{C1i}+ \rho_{C2r}+I\rho_{C2i} \right)\\ \frac{1}{2\sqrt{2}}\left( \sigma_E+\sigma_1-\sigma_2 -\sigma_3 +2\sigma_{C}\right) \end{array}\right), \\
\phi'_C&=&\left( \begin{array}{c} \frac{1}{2} \left( -\sqrt{2} H_{C} - {\bf H_1} +\pi_{1}\right)\\ f_1 + \frac{1}{2} \left( -\rho_{C1r}-I \rho_{C1i}+ \rho_{C2r}+I\rho_{C2i} \right)\\ \frac{1}{2\sqrt{2}}\left( \sigma_E+\sigma_1-\sigma_2 -\sigma_3 -2\sigma_{C}\right)\end{array} \right), \\
\phi_D&=&\left( \begin{array}{c} \frac{1}{2} \left( \sqrt{2} H_{D} - {\bf H_2} +\pi_{2}\right)\\ \frac{1}{2\sqrt{2}}\left( -\sigma^{\star}_E+\sigma^{\star}_1+\sigma^{\star}_2-\sigma^{\star}_3 +2\sigma^{\star}_{D}\right)\\ f_1 + \frac{1}{2} \left( \rho_{D1r}+I \rho_{D1i}+ \rho_{D2r}+I\rho_{D2i} \right) \end{array}\right), \\
\phi'_D&=&\left( \begin{array}{c} \frac{1}{2} \left( -\sqrt{2} H_{D} - {\bf H_2} +\pi_{2}\right)\\ \frac{1}{2\sqrt{2}}\left( -\sigma^{\star}_E+\sigma^{\star}_1+\sigma^{\star}_2 -\sigma^{\star}_3 -2\sigma^{\star}_{D}\right)\\ f_1 + \frac{1}{2} \left( -\rho_{D1r}-I \rho_{D1i}+ \rho_{D2r}+I\rho_{D2i} \right)\end{array} \right). \\
\eea
Of these fields, those in the first row are doublets under the remaining $SU(2)_W$. After plugging in the vev, the mass spectra for these doublets is
\be
2m^2 \left( |H_{A}|^2+|H_{B}|^2+
|H_{C}|^2 +|H_{D}|^2 \right).
\ee
Note that the other doublets of the theory are massless (at tree-level): $\pi_1 \ \& \ \pi_2$ are eaten by the $SU(4)$ gauge bosons; $H_1 \ \& \ H_2$ are the two light pNGB that break $SU(2)_W \times U(1)$.

The remaining singlet spectra is more complicated:
\bea
&&m^2 \left( \rho_{A1r}^2 + \rho_{B1r}^2 +  \rho_{C1r}^2 + \rho_{D1r}^2 \right) \nonumber \\
&+& \lambda v^2 \left( \rho_{A2r}^2+\rho_{B2r}^2+\rho_{C2r}^2+\rho_{D2r}^2\right)\nonumber \\
&+&\left( m^2+\lambda v^2\right) \left( \rho_{A1i}^2 + \rho_{B1i}^2 +  \rho_{C1i}^2 + \rho_{D1i}^2 \right) \nonumber \\
&+&\left(2 m^2 +4 \lambda v^2\right)\left( |\sigma_{A}|^2 + |\sigma_{B}|^2 +|\sigma_{C}|^2 + |\sigma_{D}|^2\right) \nonumber \\
&+&4 \lambda v^2 \left(2|\sigma_1|^2+|\sigma_2|^2+|\sigma_3|^2\right).
\eea
Note that the fields $\rho_{A2i}, \rho_{B2i}, \rho_{C2i},\ \& \  \rho_{D2i}$ are massless at tree level. These fields are the Goldstone bosons of the spontaneously broken $U(1)_A, U(1)_B, U(1)_C,\ \& \  U(1)_D$ symmetries. We can rewrite the fields to show more explicitly the symmetry breaking structure:
\bea
\rho_{A2i} &=&\frac{1}{2}\left( \rho_{E1}+ \rho_{E2}+\rho_3 + \rho_4 \right),\nonumber \\
\rho_{B2i} &=& \frac{1}{2}\left(\rho_{E1}- \rho_{E2}-\rho_3 + \rho_4 \right),\nonumber \\
\rho_{C2i} &=&\frac{1}{2}\left( \rho_{E1}+ \rho_{E2}-\rho_3 -\rho_4 \right),\nonumber \\
\rho_{D2i} &=&\frac{1}{2}\left( \rho_{E1}- \rho_{E2}+\rho_3 - \rho_4 \right).
\eea
The two broken combinations of the diagonal generators from $SU(4) \times U(1)_X$ eat $\rho_{E1} \ \& \ \rho_{E2}$, while the remaining fields $\rho_3, \rho_4$ receive $\sim .1 \tev$ masses from the global $U(1)$ violating term, $b_0$, described at the end of Section \ref{sec:su4}.

Hypercharge is a linear combination of one of the $SU(4)$ generators and the $U(1)_X$, under which the $\phi$ have charge $-1/4$:
\be
U(1)_Y = \frac{1}{\sqrt{2}}T^{15}+ U(1)_X.
\ee
where $T^{15} = \sqrt{2} {\rm diag} \left( -1/4, -1/4, 1/4, 1/4 \right)$ is a normalized diagonal generator of $SU(4)$ in the fundamental representation. $H_1 \  \& \ H_2$ have charges of $-1/2$ under the remaining $U(1)_Y$. The low energy gauge coupling is given by 
\be
g_Y=\frac{g g_X}{\sqrt{g^2+ \frac{g_X^2}{2}}}.
\ee

%%%%%%%%%%%%%%%%%%%%%%%%%%%%%%%%%%%%%%
\section{Quadratic Divergences} %%%%%%%%%%%%%%%%%%%%%%
\label{sec:qd} %%%%%%%%%%%%%%%%%%%%%%%%%%%%%%%

In a stage with a fundamental field $\phi$ and cutoff $\Lambda$, quadratic divergences are caused by gauge, fermion and scalar loops. 

Gauge loops contribute
\bea
\delta V_{gb} &=&  \frac{3 g^2 \Lambda^2}{16 \pi^2} C_2(\phi) \phi^{\dagger} \phi^{},
\eea
where we use the normalization ${\rm Tr}\left( T^a T^b \right)=1/2$ throughout.

A single Yukawa interaction of the form $y \phi^{\dagger} \hat{Q} T^c_{A}$ gives a quadratic divergence

\bea
\delta V_{f} &=& -\frac{N_c \Lambda^2}{8 \pi^2} y^{2} \phi^{\dagger}\phi.
\eea

while the doubled Yukawa interaction of Section \ref{sec:u1z}, $\left(y \chi + {y'} {\chi'}\right){\tilde{Q}}{tilde{T}}$ gives

\bea
\delta V_{f} &=& -\frac{N_c \Lambda^2}{8 \pi^2} \left(y^{} \chi^{\dagger}+{y'}^{}{\chi'}^{\dagger}\right)\left(y^{}\chi+{y'}^{}{\chi'}\right). 
\eea

A scalar quartic $\lambda (\phi^{\dag} {\phi'}^{})({\phi'}^{\dagger} \phi^{})$ gives

\bea
\delta V_{s} &=& \frac{\lambda \Lambda^2}{16 \pi^2} \left( \phi^{\dagger} {\phi'}^{}+{\phi'}^{\dagger} \phi^{} \right).
\eea

%%%%%%%%%%%%%%%%%%%%%%%%%%%%%%%%%%%%%%
\section{RGE's} %%%%%%%%%%%%%%%%%%%%%%%%%%%%%%
\label{sec:RGE} %%%%%%%%%%%%%%%%%%%%%%%%%%%%%%

The gauge coupling runs as
\bea
\frac{d \alpha^{-1}}{dt}&=&\frac{b_0}{2 \pi}.
\eea
  For the $SU(2) \times U(1)$ theory, we have $(b_Y, b_W, b_c)=(-7, 3, 7)$, for the $SU(4) \times U(1)$ theory we have $(b_X, b_W, b_c)=(-32/3, 28/3, 3)$, while for the $SU(12) \times U(1)$ theory we have $(b_Z, b_W, b_c) = (-278/9, 104/3, -13)$.

For the $SU(4) \times U(1)_X$ sector, considering just the 3rd generation couplings
\be V \supset \left\{ y^{}_{TA} \phi_A^{\dagger}T^c_A +y^{}_{TB}{\phi}_B^{\dagger}{T}^c_B + y^{}_{TC} \phi_C^{\dagger}T^c_C \right\} \hat{Q_3}, \ee
we find the typical Yukawa runs as 
\bea
\frac{d y_{TA}}{dt}&=&\frac{1}{16 \pi^2} y_{TA} \left\{ \frac{1}{2} \left( y_{TA}^2 + y_{TB}^2 + y_{TC}^2\right)+ \frac{1}{2}\left(N_W y_{TA}^2\right) + N_C y_{TA}^2 \right. \nonumber \\
&& \hspace{.5 in} \left. - 6 g_c^2 \frac{N_c^2 -1}{2 N_C} - 3 g_W^2 \frac{N_W^2 -1}{2 N_W} - 3 g_x^2 \left( x_Q^2 + x_T^2 \right) \right\}.
\eea
In the $SU(12) \times U(1)_Z$ case, every factor within the brackets of $y_{TX}^2 \rightarrow y_{TX}^2 + {y'}_{TX}^2$ and the the first sum in the parentheses must sum over all Yukawa coefficients for the third generation quark.

We assume that the coefficients in the scalar sector are
properly symmetrized where applicable, e.g. $\rho_{ij} =
\rho_{ji}$ and where $i,j \in \{A, A', B, B', \ldots \}$. 

\bea
V &=& \frac{1}{2}\sum_{i,j:i \ne j} \rho_{ij} \left( \phi_i^{\dag} \phi_j\right)\left( \phi_j^{\dag} \phi_i\right),
\eea
with RGE's (don't sum over repeated indices, except where explicitly indicated)

\bea
\frac{d \rho_{ij}}{dt}&=&\frac{1}{8 \pi^2}\left[N \rho_{ij}^2 + \sum_{k \neq i,j} \rho_{ik}\rho_{kj} - \frac{3 (N^2-1)}{N} \rho_{ij} g^2 -6 g_x^2 \left( X_Q^2 + X_T^2\right)\right. \\ \nonumber 
&&\hspace{.5 in}\left.+\frac{3 (N^2-4)}{4N} g^4+ \rho_{ij} \left( y_{Ti}^2 + y_{Tj}^2\right) -2 N_c y_{Ti}^2 y_{Tj}^2\right].
\eea
In the $SU(12) \times U(1)_Z$ sector, the final factor of $-2 N_c \rightarrow -4 N_c$ if $i,j$ are a pair like $\{(A, A'), (B, B'), \ldots\}$.

\end{appendix}


\begin{thebibliography}{99}
\bibitem{mt} However, we shall use the name ``melba toast'' interchangeably.
\end{thebibliography}
\end{document}